%% file: acl_latex.tex
\documentclass[11pt]{article}

\usepackage[final]{acl}

\usepackage{times}
\usepackage{latexsym}
\usepackage{booktabs}
\usepackage{multirow}
\usepackage{CJKutf8}
\usepackage{amssymb}
\usepackage[T1]{fontenc}

\usepackage[utf8]{inputenc}

\usepackage{microtype}

\usepackage{inconsolata}

\usepackage{graphicx}
\usepackage{array}    
\usepackage{xspace}
\usepackage{amsmath}
\usepackage{url}
\usepackage{subcaption}
\title{Preference Shapes Relevance: Cross-component Hierarchical Semantic Alignment for Personalized Generative Retrieval}

\author{
  \textbf{Gaoming Zhang}\textsuperscript{1} \quad
  \textbf{Angqing Jiang}\textsuperscript{1} \quad
  \textbf{Jianchun Song}\textsuperscript{2} \quad
  \textbf{Kena Qi}\textsuperscript{2} \\
  \textbf{Dayao Chen}\textsuperscript{2} \quad
  \textbf{Wei Lin}\textsuperscript{2} \quad
  \textbf{Defu Lian}\textsuperscript{1}\thanks{Corresponding author.} \\
  \textsuperscript{1}University of Science and Technology of China \\
  \textsuperscript{2}Meituan \\
  \texttt{\{zzzgm, philipgaq\}@mail.ustc.edu.cn} \quad
  \texttt{liandefu@ustc.edu.cn} \\
  \texttt{\{songjianchun, qikena, chendayao, linwei31\}@meituan.com}
}

\begin{document}
\maketitle
\begin{abstract}
  \input{Sections/0_Abstract.tex}
\end{abstract}

\section{Introduction}
\input{Sections/1_Introductions.tex}

\section{Related Works}
\input{Sections/2_Related_Works.tex}

\section{Preliminaries}

\input{Sections/3_Preliminaries.tex}

\section{Methodology}

\input{Sections/4_Methodology.tex}

\section{Experiments}
\input{Sections/5_Experiments.tex}

\section{Conclusion}
\input{Sections/6_Conclusion.tex}

\section*{Limitations}
\input{Sections/7_Limitations.tex}

\section*{Ethical Considerations}
\input{Sections/8_Ethical_Considerations.tex}

\section*{Acknowledgements}
The work is supported by grants from New Generation Artificial Intelligence-National Science and Technology Major Project (2025ZD0123403), National Natural Science Foundation of China (No. U24A20253), Scientific Research Innovation Capability Support Project for Young Faculty, and the Meituan Research Fund.

\bibliography{Sections/References.bib}

\newpage
\appendix
\input{Sections/Appendix.tex}

\end{document}

%% file: Sections/0_Abstract.tex
Generative Retrieval (GR) has emerged as a promising paradigm by mapping queries directly to Semantic IDs (SIDs) with powerful representation capabilities for candidate items. However, existing SIDs derived solely from item content create a semantic gap, failing to align dynamic query intents with static item representations. Furthermore, current generative paradigms rarely model user behavior sequences and are always bottlenecked by the high inference
latency of beam-search autoregressive decoding. To address these challenges, we propose \textbf{C}ross-component \textbf{H}ierarchical semantic \textbf{A}lignment for \textbf{P}ersonalized generative retrieval (\textbf{CHAP}), a novel personalized GR framework from a hierarchical perspective. First, we design a Hierarchical Semantic Alignment module to align query's latent space with item's quantization path and synchronize multi-granular semantics. Second, we construct a personalized GR framework that models user behavior by synergizing discrete SIDs for structural guidance and continuous representations for fine-grained semantic refinement. Notably, we introduce a Residual Cascading Generation mechanism to restrict the costly multi-step Transformer Decoder to a single-pass inference, boosting inference throughput while mitigating information loss. Extensive experiments on three public datasets, one proprietary industrial dataset, and online A/B tests demonstrate CHAP's superiority, validating the effectiveness and practical value of our approach. The code is publicly available at \textcolor{blue}{\url{https://github.com/zzzgm/CHAP}}.

%% file: Sections/1_Introductions.tex
Modern Information Retrieval (IR) systems traditionally rely on the "Index-Retrieve-Rank" cascade pipeline~\cite{wang2011cascade,zhou2021survey}, employing sparse methods like BM25~\cite{robertson2009probabilistic} for keyword matching and dense methods like DPR~\cite{karpukhin2020dense} for semantic vector mapping. However, static models often struggle to adapt to dynamic, personalized user intents~\cite{liu2020personalization}.
Recently, Generative Retrieval (GR) has emerged as a disruptive paradigm alongside the advancement of Large Language Models (LLMs)~\cite{tang2023recent,li2025matching}. By adopting an end-to-end approach that transforms retrieval into a sequence generation task which directly generates relevant item identifiers (ItemIDs) from user queries, GR eliminates the need for complex external indices. This paradigm significantly enhances scalability and demonstrates superior performance in long-tail queries and cold-start problems~\cite{kuo2024survey,yuan2024generative}.

In the GR architecture, generating Semantic IDs (SIDs) with robust representational capabilities remains a core challenge~\cite{zhang2025multi,liu2025cat}. Existing models like TIGER~\cite{rajput2023recommender} and HierGR~\cite{zhang2025hiergr} train SIDs solely on item corpora, lacking explicit query perception and failing to bridge the inherent \textit{semantic gap} between user intents and item descriptions, which leads to poor generalization on complex queries~\cite{zhang2025c2t,zheng2025enhancing}. Furthermore, current methods exhibit critical limitations in structural utilization, personalization, and inference efficiency. Treating quantized SIDs merely as flat symbol sequences ignores the natural "Coarse-to-Fine" hierarchy, wasting its benefits for feature generalization and semantic stability~\cite{williams2020hierarchical,si2024generative}. Relying solely on discrete sparse IDs also discards fine-grained continuous vector details, hindering the integration of user behavior sequences~\cite{yang2025sparse}. Crucially, standard autoregressive GR paradigms suffer from severe computational bottlenecks during online serving; repeatedly executing the heavy Transformer Decoder (especially the costly Cross-Attention) for each hierarchical step causes prohibitive latency, hindering large-scale deployment in latency-sensitive industrial systems.

To address these limitations, we propose \textbf{C}ross-component \textbf{H}ier\-archical semantic \textbf{A}lignment for \textbf{P}ersonalized generative retrieval (\textbf{CHAP}), a novel framework which performs hierarchical alignment on pre-trained SIDs and efficiently fuses sparse and dense representations in a sequence model.

Specifically, CHAP initializes SIDs using a Residual Quantized Variational Autoencoder (RQ-VAE)~\cite{rajput2023recommender}. To bridge the query-item semantic gap while preventing codebook collapse, we design a multi-module \textbf{Hierarchical Semantic Alignment} paradigm which includes a cross-sample alignment module to proactively pull the query's potential representation towards the target's quantization space, a hierarchical-aware contrastive learning module to enforce coarse-to-fine alignment, and a soft probability distillation module to regularize the output distribution of the encoder via KL divergence. Integrating these modules ensures the generated SIDs can robustly comprehend complex query semantics.

Furthermore, CHAP constructs a \textbf{Personalized Cascading Generative Sequence Model} that implements multi-granular modeling of user queries and personalized behavior sequences by cross-cascading aligned sparse SIDs with its raw dense vector. Unlike standard autoregressive approaches, CHAP implements a \textbf{Residual Cascading Generation} strategy which decouples heavy historical intent routing from hierarchical code generation. By restricting the costly multi-step Transformer Decoder to a single-pass inference and offloading the layer-wise generation to lightweight residual blocks, this mechanism not only mitigates information loss but also acts as an efficient decoding accelerator, significantly boosting inference throughput.
Our main contributions are as follows:
\begin{itemize}
    \item[$\bullet$] We propose a Hierarchical Semantic Alignment paradigm, effectively bridging the inherent Semantic Gap between user intents and static item codes.
    \item[$\bullet$] We design a Personalized GR Architecture equipped with a Residual Cascading Generation mechanism, which resolves the inherent latency bottleneck of autoregressive GR while effectively mitigating information loss through coarse-to-fine structural guidance.
    \item[$\bullet$] Extensive experiments on three public datasets and one proprietary industrial dataset, as well as online A/B tests demonstrate the superiority of our method.
\end{itemize}

%% file: Sections/2_Related_Works.tex
\textbf{Sparse retrieval} methods like BM25~\cite{robertson2009probabilistic} and TF-IDF~\cite{ramos2003using} rely on exact term matching but suffer from vocabulary mismatch. Conversely, \textbf{dense retrieval} approaches (e.g., DPR~\cite{karpukhin2020dense}, ANCE~\cite{xiong2020approximate}) map inputs into a shared vector space, enhancing semantic matching but necessitating costly external indices like HNSW~\cite{malkov2018efficient}. \textbf{Personalized retrieval} tailors results by incorporating user contexts. QEM~\cite{ai2019zero} investigates personalization mechanisms, while DREM~\cite{ai2019explainable} models dynamic user-product relationships via knowledge graphs. Embedding-based approaches like ZAM~\cite{ai2019zero}, HEM~\cite{ai2017learning}, and TEM~\cite{bi2020transformer} integrate user profiles and history into unified embeddings.

\textbf{Generative retrieval} internalizes corpus knowledge to directly generate ItemIDs. DSI~\cite{tay2022transformer} pioneered this paradigm, followed by semantic enhancements in SE-DSI~\cite{tang2023semantic}. TIGER~\cite{rajput2023recommender} first utilized RQ-VAE for SIDs. Regarding architecture, COBRA~\cite{yang2025sparse} integrates sparse and dense vectors, while CAT-ID$^2$~\cite{liu2025cat}, and MERGE~\cite{zhang2025multi} optimize ID generation by exploiting multi-level correlations.

%% file: Sections/3_Preliminaries.tex
\subsection{Problem Definition}
Given a user's interaction history $S = \{i_1, \dots, i_N\}$ and a current query $q$, Personalized GR aims to retrieve items by directly generating their identifiers such as a unique hierarchical SID $\mathbf{c} = (c^0, \dots, c^{L-1})$ derived from a codebook. 
The objective is to learn a probabilistic model $\theta$ that maximizes the probability of generating the target SID given the context:
\begin{equation}
    \begin{aligned}
    P(i | q, S; \theta) &= P(\mathbf{c} | q, S; \theta) \\
    &= \prod_{l=0}^{L-1} P(c^l | c^{<l}, q, S; \theta).
    \end{aligned}
\end{equation}
where $c^{<l}$ denotes the generated prefix tokens at layer $l$. Inference is to generate the target SID with the highest joint probability.

\begin{figure}[t] 
    \centering
    \includegraphics[width=0.48\textwidth]{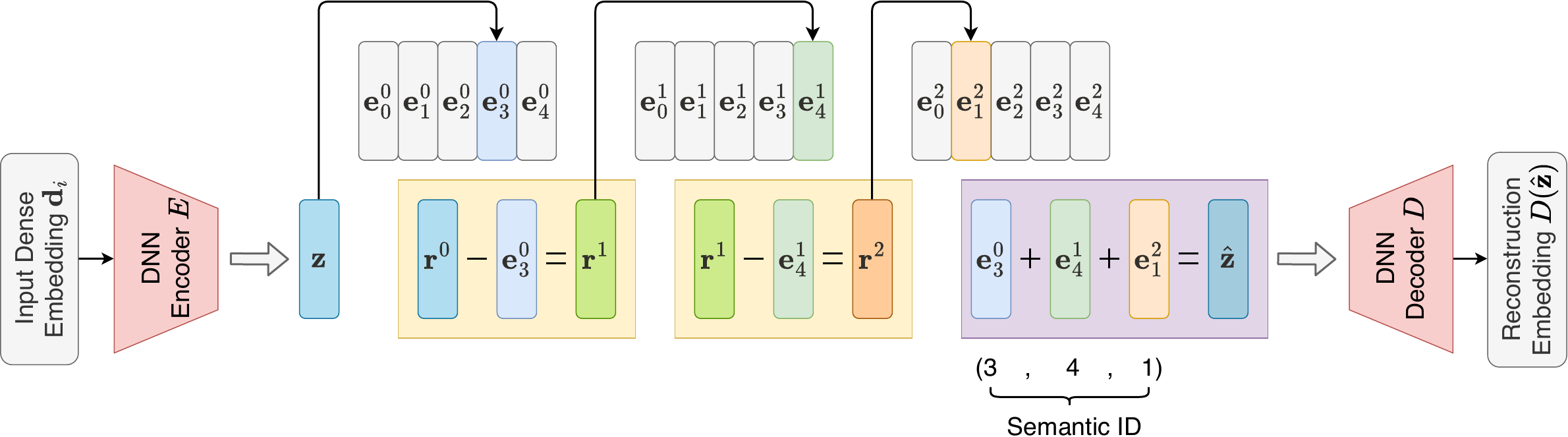}
    \caption{Schematic of SID generation via RQ-VAE.}
    \label{fig:rqvae}
\end{figure}

\subsection{Semantic ID Generation via RQ-VAE}
\label{sec:rqvae}
To assign meaningful discrete identifiers to items effectively, we employ RQ-VAE~\cite{rajput2023recommender} (Fig.~\ref{fig:rqvae}). This approach generates hierarchical SIDs by recursively quantizing item embeddings in a coarse-to-fine manner, which serves as the foundational representation for our proposed framework.

Given an item $i$, a pre-trained language encoder first encodes its textual content into a dense embedding $\mathbf{d}_i$. A deep neural network encoder $E(\cdot)$ then maps this embedding to a latent representation $\mathbf{z} = E(\mathbf{d}_i)$. The quantization process utilizes $L$ residual layers, where each layer $l$ is equipped with a codebook $\mathcal{C}^{l}=\{\mathbf{e}_{k}^{l}\}_{k=1}^{K}$ containing $K$ code words. 
At each step $l$, quantization is performed by selecting the code word nearest to the current residual vector $\mathbf{r}^l$ (where $\mathbf{r}^0 = \mathbf{z}$). The \textit{Semantic Token} $c^l$ for layer $l$ is determined as:
\begin{equation}
    c^{l} = \underset{k}{\arg\min} \| \mathbf{r}^{l} - \mathbf{e}_{k}^{l} \|_{2}^{2}, \quad \mathbf{r}^{l+1} = \mathbf{r}^{l} - \mathbf{e}_{c^{l}}^{l}.
\end{equation}

After $L$ iterations, we obtain a discrete coarse-to-fine SID $(c^{0}, \dots, c^{L-1})$. The quantized latent representation $\hat{\mathbf{z}}$ is then approximated by the sum of the selected codebook vectors: $\hat{\mathbf{z}} = \sum_{l=0}^{L-1} \mathbf{e}_{c^{l}}^{l}$.

The training of RQ-VAE aims to reconstruct the original input while optimizing the codebooks. The decoder $D(\cdot)$ attempts to reconstruct the input $\mathbf{d}_i$ from $\hat{\mathbf{z}}$, resulting in the reconstruction loss:
\begin{equation}
    \mathcal{L}_{\text{recon}} = \| \mathbf{d}_i - D(\hat{\mathbf{z}}) \|_{2}^{2}.
\end{equation}

To ensure the stability of quantization, RQ-VAE employs a split quantization objective consisting of a codebook loss to update codebook vectors towards the residuals and a commitment loss to constrain the encoder output:
\begin{align}
    \mathcal{L}_{\text{codebook}} &= \sum_{l=0}^{L-1} \| sg[\mathbf{r}^{l}] - \mathbf{e}_{c^{l}}^{l} \|_{2}^{2} \\
    \mathcal{L}_{\text{commit}} &= \beta \sum_{l=0}^{L-1} \| \mathbf{r}^{l} - sg[\mathbf{e}_{c^{l}}^{l}] \|_{2}^{2},
\end{align}
where $sg[\cdot]$ denotes the stop-gradient operation which prevents gradient update during backpropagation, and $\beta$ is a hyperparameter balancing the commitment loss. The final training objective is:
\begin{equation}
    \mathcal{L}_{\text{RQ-VAE}} = \mathcal{L}_{\text{recon}} + \mathcal{L}_{\text{codebook}} + \mathcal{L}_{\text{commit}}.
\end{equation}

%% file: Sections/4_Methodology.tex
\subsection{Hierarchical Semantic Alignment (HSA)}
\begin{figure*}[t]
    \centering
    \includegraphics[width=0.98\textwidth]{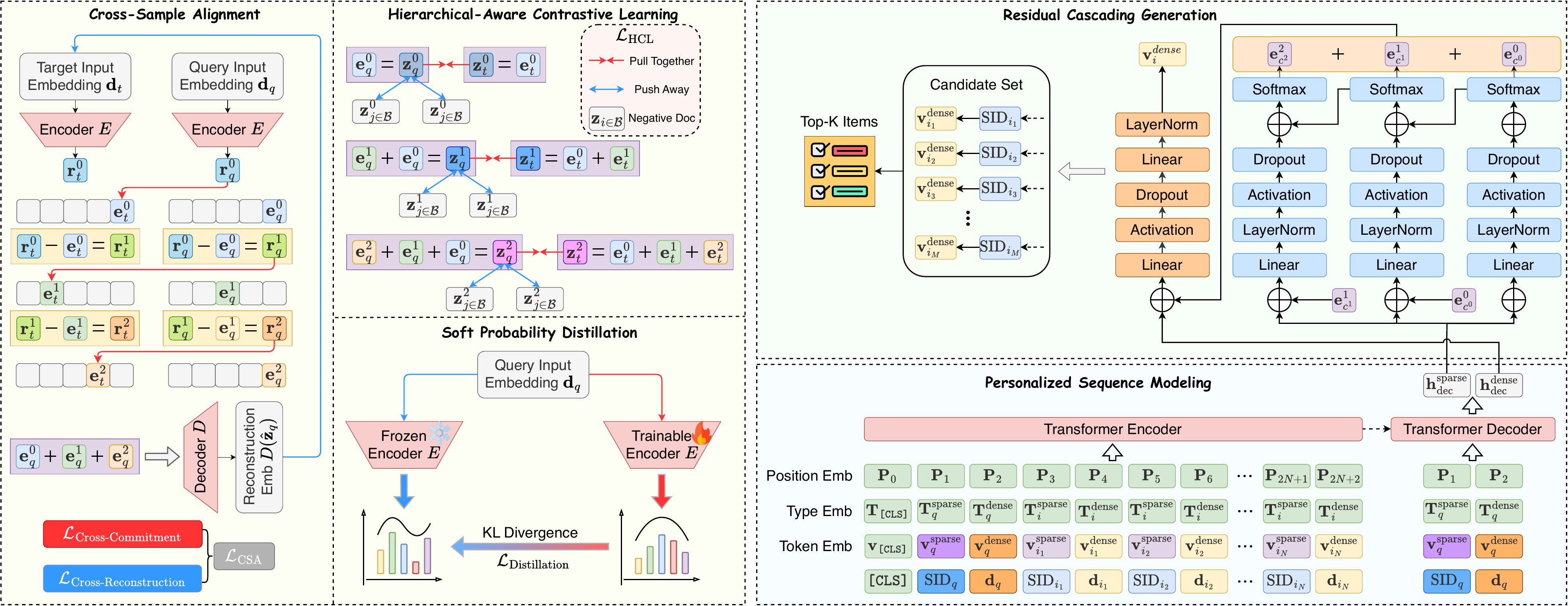}
    \caption{\textbf{Overall architecture of CHAP}, featuring HSA to bridge the query-item gap (left), and dual-view Personalized Sequence Modeling with Residual Cascading Generation for efficient coarse-to-fine retrieval (right).}
    \label{fig:hsa}
\end{figure*}

Directly applying an item-pretrained RQ-VAE to retrieval faces a semantic gap, as its encoder $E(\cdot)$ is optimized solely for item reconstruction. To bridge this, we propose a Hierarchical Semantic Alignment (HSA) paradigm (Fig.~\ref{fig:hsa}). To prevent semantic drift and catastrophic forgetting, we freeze the pre-trained codebook, forcing the query quantization path directly into the stable item manifold.

\textbf{Cross-Sample Alignment (CSA).} To pull the query's latent representation into the item's quantization space, we repurpose RQ-VAE losses by swapping input-target variables. Given a query $q$ and target $t$, the \textit{Cross-Commitment} loss forces the query's residual $\mathbf{r}_q^l$ to align with the target's selected code $\mathbf{e}_{c^l}^l$:
\begin{equation}
    \mathcal{L}_{\text{CC}} = \beta \sum_{l=0}^{L-1} \| \mathbf{r}_q^l - sg[\mathbf{e}_{c^l}^l] \|_{2}^{2}.
\end{equation}
The \textit{Cross-Reconstruction} loss ensures the target's raw content $\mathbf{d}_t$ is reconstructable from the query's quantized representation $\hat{\mathbf{z}}_q$:
\begin{equation}
    \mathcal{L}_{\text{CR}} = \| \mathbf{d}_t - D(\hat{\mathbf{z}}_q) \|_{2}^{2}.
\end{equation}
The total CSA loss is $\mathcal{L}_{\mathrm{CSA}} = \mathcal{L}_{\mathrm{CC}} + \mathcal{L}_{\mathrm{CR}}$.

\textbf{Hierarchical-Aware Contrastive Learning (HCL).} To explicitly exploit the ``Coarse-to-Fine'' SID structure, we align the query's partial quantized representation $\hat{\mathbf{z}}_q^{(l)} = \sum_{i=0}^{l} \mathbf{e}_{c^i_q}^i$ with the target's partial quantized representation $\hat{\mathbf{z}}_t^{(l)} = \sum_{i=0}^{l} \mathbf{e}_{c^i_t}^i$ accumulated up to depth $l$:
\begin{equation}
\label{eq:hcl}
    \begin{aligned}
    \mathcal{L}_{\text{HCL}}
    &= -\sum_{l=0}^{L-1} \log
    \frac{\exp(s_l^+/\tau)}{\sum_{j \in \mathcal{B}} \exp(s_{lj}/\tau)}, \\
    s_l^+ &= \operatorname{sim}\!\left(\hat{\mathbf{z}}_q^{(l)},
    sg[\hat{\mathbf{z}}_t^{(l)}]\right), \\
    s_{lj} &= \operatorname{sim}\!\left(\hat{\mathbf{z}}_q^{(l)},
    sg[\hat{\mathbf{z}}_j^{(l)}]\right).
    \end{aligned}
\end{equation}
where $\mathcal{B}$ is the batch containing the positive target and negatives, and $\tau$ is the temperature. 

\textbf{Soft Probability Distillation.} To prevent drastic latent shifts caused by hard quantization assignments, we distill the soft code assignment distribution from the frozen item-side quantization path of the target item to the trainable query encoder via KL divergence:
\begin{equation}
    \mathcal{L}_{\text{Distill}} = \sum_{l=0}^{L-1} \text{KL}\left( P_{\text{item}}(\cdot|t) \parallel P_{\text{query}}(\cdot|q) \right),
\end{equation}
where the soft assignment probability is defined by the Euclidean distance to the residual:
\begin{equation}
    P(c^l=k|q) = \frac{\exp\left(- \| \mathbf{r}_q^l - \mathbf{e}_{k}^{l} \|_2^2\right)}{\sum_{j=1}^{K} \exp\left(- \| \mathbf{r}_q^l - \mathbf{e}_{j}^{l} \|_2^2\right)}.
\end{equation}
The comprehensive HSA objective is optimized via a weighted sum:
\begin{equation}
    \mathcal{L}_{\mathrm{HSA}} = \mathcal{L}_{\mathrm{CSA}} + \gamma \cdot \mathcal{L}_{\mathrm{HCL}} + \delta \cdot \mathcal{L}_{\mathrm{Distill}}.
\end{equation}

\subsection{Personalized Sequence Modeling}
\textbf{Dual-View Sequence Modeling.} To capture the multi-granularity preferences of users and generate high-fidelity retrieval representations, we construct an input sequence where each interaction synergizes an aligned discrete SID and a continuous dense vector. For a user history $S = \{i_1, \dots, i_N\}$ and query $q$, the input is organized as:
\begin{equation}
    \begin{aligned}
    X_{in} = [&\mathbf{v}_{\texttt{[CLS]}}, \mathbf{v}_q^\mathrm{sparse},
    \mathbf{v}_q^\mathrm{dense}, \\
    &\mathbf{v}_{i_1}^\mathrm{sparse}, \mathbf{v}_{i_1}^\mathrm{dense}, \dots].
    \end{aligned}
\end{equation}
The sparse view $\mathbf{v}^\mathrm{sparse}$ maps the aggregated codebook vectors ($\sum \mathbf{e}_{c^l}^l$) into a unified dense space, while $\mathbf{v}^\mathrm{dense}$ corresponds to the raw language model embedding. To maintain training-inference consistency and prevent exposure bias, the unified decoder is initialized exclusively with the query's dual representation ($\mathbf{v}_q^\mathrm{sparse}, \mathbf{v}_q^\mathrm{dense}$). This explicitly forces the decoder to ``look back'' via cross-attention to retrieve personalized results.

\textbf{Hybrid Optimization.} Our multi-task framework optimizes the \textbf{Sparse Generation Loss} for hierarchical SID classification across all $L$ levels:
\begin{equation}
    \mathcal{L}_\mathrm{sparse} = - \sum_{l=0}^{L-1} \log \left( \frac{\exp({s}^l_{c^l})}{\sum_{k=1}^{K} \exp({s}^l_{k})} \right),
\end{equation}
where $\mathbf{s}^l \in \mathbb{R}^{K}$ is the logit vector at level $l$. Simultaneously, to avoid Mean Squared Error instability, we employ an InfoNCE-based \textbf{Dense Contrastive Loss} ($\mathcal{L}_\mathrm{dense}$) over the batch set $\mathcal{B} = \{\mathbf{v}^+\} \cup \mathcal{B}^-$:
\begin{equation}
    \mathcal{L}_\mathrm{dense} = - \log \frac{\exp(\text{sim}(\hat{\mathbf{v}}, \mathbf{v}^+) / \tau)}{\sum_{\mathbf{v}' \in \mathcal{B}} \exp(\text{sim}(\hat{\mathbf{v}}, \mathbf{v}') / \tau)},
\end{equation}
treating the predicted dense vector $\hat{\mathbf{v}}$ (introduced in Sec.~\ref{subsec:residual_cascading_generation}) as the anchor. The joint objective $\mathcal{L}_\mathrm{total} = \mathcal{L}_\mathrm{sparse} + \mathcal{L}_\mathrm{dense}$ creates a robust curriculum learning effect, where sparse structural constraints simplify the dense ranking optimization.

\subsection{Residual Cascading Generation and Retrieval}
\label{subsec:residual_cascading_generation}
To achieve a balance between structural correctness and fine-grained semantic precision, we adopt a ``Coarse-to-Fine'' strategy to coordinate the prediction of discrete SIDs and continuous dense vectors. Formally, we decompose the generation probability of a target item $i$, given the context $X$, into two stages: the joint probability of the $L$ hierarchical Semantic Tokens, followed by the conditional probability of the dense vector (operationalized by a cosine-softmax score over generated candidates):
\begin{equation}
    P(i|X) =
    \underbrace{\prod_{l=0}^{L-1} P(c_i^l | c_i^{<l}, X)}_{\text{Sparse Generation}}
    \cdot
    \underbrace{P(\mathbf{v}_i | c_i^{0:L-1}, X)}_{\text{Dense Refinement}}.    
\end{equation}
An entropy-based justification for this cascaded sparse-dense factorization is
provided in App.~\ref{app:cascade_proof}.

\textbf{Coarse-to-Fine Residual Generation.} To avoid the prohibitive latency of standard autoregressive decoding, we propose a Single-pass Residual Generation Mechanism. The Transformer Decoder executes only once over the query's input tokens. Specifically, let $\mathbf{h}_{\mathrm{dec}}^{\mathrm{sparse}}$ and $\mathbf{h}_{\mathrm{dec}}^{\mathrm{dense}}$ denote the output hidden states corresponding to the discrete query SID position and the continuous query embedding position, respectively. Utilizing $\mathbf{h}_{\mathrm{dec}}^{\mathrm{sparse}}$ as a global structural context anchor, subsequent states for predicting level $l$ are updated via lightweight residual blocks:
\begin{equation}
    \mathbf{h}^{(l+1)} = \operatorname{ResBlock} \left( \text{Concat}[ \mathbf{h}_{\mathrm{dec}}^{\mathrm{sparse}}, \mathbf{e}_{c^l}^l ] \right) + \mathbf{h}^{(l)}.
\end{equation}
This decoupled design acts as a powerful decoding accelerator. To align the continuous vector with this discrete structure, we inject the reconstructed quantization $\hat{\mathbf{z}} = \sum_{l=0}^{L-1} \mathbf{e}_{c^l}^l$ into the dense prediction head conditioned on $\mathbf{h}_{\mathrm{dec}}^{\mathrm{dense}}$:
\begin{equation}
    \hat{\mathbf{v}} = \operatorname{PredHead} \left( \text{Concat}[ \mathbf{h}_{\mathrm{dec}}^{\mathrm{dense}}, \hat{\mathbf{z}} ] \right),
\end{equation}
achieving an elegant balance between structural correctness and fine-grained semantic precision.

\textbf{Candidate Scoring.} During inference, we maximize throughput via a \textbf{parallel sampling strategy} to generate $M$ candidate SIDs. We calculate the accumulated log-probability of the sampled hierarchical path as the sparse signal $S_\mathrm{sparse}$, and the cosine similarity between the predicted dense vector $\hat{\mathbf{v}}$ corresponding to the sampled SID and the candidate item’s raw vector $\mathbf{v_i}$ as the dense signal $S_\mathrm{dense}$. Both signals are normalized using a temperature-scaled Softmax over the candidate pool $\mathcal{I}_\mathrm{cand}$:
\begin{equation}
    \mathcal{N}(S) = \frac{\exp(S / \tau)}{\sum_{j \in \mathcal{I}_\mathrm{cand}} \exp(S_j / \tau)}.
\end{equation}
The final score fuses both dual-view intents:
\begin{equation}
    S_\mathrm{final}(i) = \mathcal{N}\left( S_\mathrm{sparse}(i) \right) \cdot \mathcal{N}\left( S_\mathrm{dense}(i) \right).
\end{equation}

%% file: Sections/5_Experiments.tex
\input{Sections/Tables/dataset_stat.tex}

We analyze CHAP's effectiveness and address these research questions in this section: \textbf{RQ1:} How does CHAP perform against state-of-the-art baselines? \textbf{RQ2:} How does semantic alignment improve generated ItemID quality? \textbf{RQ3:} How do core components contribute across intent-wise query segments? \textbf{RQ4:} Does our residual architecture resolve GR latency bottlenecks? \textbf{RQ5:} How robust is CHAP to codebook and decoding hyperparameters? \textbf{RQ6:} Can CHAP drive tangible business growth in live online production?

\subsection{Experimental Settings}

\subsubsection{Datasets} 
To validate the effectiveness of CHAP, we conduct extensive experiments on four real-world datasets. \textbf{ESCI-us}~\cite{reddy2022shopping} features authentic Amazon search queries with complex linguistic patterns and strict attribute constraints. \textbf{KuaiSearch}~\cite{li2026kuaisearch} provides a massive e-commerce search corpus from Kuaishou, preserving real queries, long-term user behavior sequences, and long-tail product distributions. \textbf{Amazon}~\cite{cai2025large} integrates diverse user profiles and historical web behaviors to rigorously evaluate highly personalized search instructions. Lastly, \textbf{Local-Life} is a large-scale industrial dataset collected from 30 days of user interaction logs on a leading platform, covering heterogeneous search intents with raw natural language texts. Detailed descriptions and comprehensive statistics for all four datasets are provided in Tab.~\ref{tab:dataset_stat} and App.~\ref{app:dataset_details}.

\input{Sections/Tables/overall_result.tex}

\input{Sections/Tables/ablation.tex}

\subsubsection{Baselines}
To comprehensively evaluate the performance of CHAP, we compare it against a diverse set of 14 representative baselines across four categories. For sparse retrieval, we select \textbf{BM25}~\cite{robertson2009probabilistic}, \textbf{Doc2Query}~\cite{nogueira2019document}, and \textbf{DeepCT}~\cite{dai2019context}. For dense retrieval, we employ \textbf{DPR}~\cite{karpukhin2020dense}, \textbf{Sen-T5}~\cite{ni2022sentence}, and \textbf{MPNet}~\cite{song2020mpnet}. For personalized retrieval scenarios, we compare against \textbf{TEM}~\cite{bi2020transformer} and \textbf{CoPPS}~\cite{dai2023contrastive}. Finally, for Generative Retrieval (GR), we evaluate against state-of-the-art models including \textbf{DSI}~\cite{tay2022transformer}, \textbf{NCI}~\cite{wang2022neural}, \textbf{TIGER}~\cite{rajput2023recommender}, \textbf{LTRGR}~\cite{li2024learning}, \textbf{MERGE}~\cite{zhang2025multi}, and \textbf{COBRA}~\cite{yang2025sparse}. Detailed descriptions are provided in App.~\ref{app:baselines}.

\subsubsection{Evaluation Metrics}
We evaluate retrieval performance using four standard metrics: \textbf{Recall} (R@K) for the overall macroscopic coverage of relevant candidates; \textbf{Hit Ratio} (HR@K) for the success rate of retrieving at least one target in sparse or single-interaction scenarios; \textbf{Mean Reciprocal Rank} (MRR@K) for assessing strict positional ranking quality in single-target datasets; and \textbf{Normalized Discounted Cumulative Gain} (NDCG@K) for handling fine-grained permutations in multi-target datasets. Detailed mathematical formulations and selection justifications are provided in App.~\ref{app:metrics}.

\subsubsection{Implementation Details}
We utilize BERT-base~\cite{devlin2019bert} for raw text representation and T5-base~\cite{raffel2020exploring} for the generative retrieval backbone (Chinese-BERT-base and mT5-base for Chinese datasets). In the SID learning stage, the RQ-VAE is configured with a codebook depth of $L=3$ and a codebook size of $K=512$. For the Hierarchical Semantic Alignment, the loss weights $\gamma$ and $\delta$ are set to $0.01$ and $0.001$, respectively. During inference, we employ a parallel sampling strategy to efficiently generate $M=50$ diverse candidate SIDs, and set the temperature $\tau=1.1$ for the joint probability normalization operator $\mathcal{N}(\cdot)$. Comprehensive details and explicit implementation settings for all baseline models are provided in App.~\ref{app:implementation}.

\subsection{Overall Results (RQ1)}
Tab.~\ref{tab:main_results} presents the comparative results across the four evaluated datasets. CHAP consistently establishes a new state-of-the-art, significantly outperforming strong baselines from all categories, including advanced GR models like COBRA and MERGE. Specifically, while standard GR methods like TIGER struggle with the inherent semantic gap—often lagging behind traditional BM25, our Hierarchical Semantic Alignment effectively bridges this gap, synergizing the semantic comprehension of dense retrievers with the structural precision of GR. Furthermore, traditional personalized baselines frequently suffer from relevance drift, allowing historical behaviors to overshadow current search intents. In contrast, CHAP dynamically refines query relevance through its dual-view sequence modeling. This prevents relevance drift and ensures retrieved items are precisely aligned with both the explicit query constraints and the user's fine-grained intents, yielding robust generalization across diverse, real-world search scenarios.

Notably, unlike COBRA~\cite{yang2025sparse}, which mandates the generation of multiple dense representations—incurring excessive training costs and prohibitive real-time inference overhead, CHAP achieves superior efficiency by elegantly reusing the item's raw representations for sequence modeling. Furthermore, MERGE~\cite{zhang2025multi} heavily relies on aligning multiple positive samples and varying relevance levels under a single query, exhibiting limited generalization on datasets with single positive targets or uniform relevance levels. In contrast, our semantic alignment paradigm is specifically designed to bridge the gap between complex, multi-intent queries and targets.

\subsection{Ablation Study}
Tab.~\ref{tab:ablation} details our ablation study on Local-Life, incorporating Queries Per Second (QPS) to concurrently assess industrial efficiency. Detailed variant configurations are provided in App.~\ref{app:ablation_details}.

\subsubsection{Analysis of Generated SIDs (RQ2)}
Replacing our aligned encoder with a fixed pre-trained encoder (w/o HSA) causes significant degradation, confirming the critical necessity of proactive alignment in bridging the semantic gap. Furthermore, we observe a performance drop when removing the continuous vectors (w/o Dense Refinement). This validates our dual-view design intuition: sparse SIDs act strictly as structural skeletons for coarse retrieval, while dense vectors are indispensable for carrying the fine-grained semantics required for precise personalized matching.

\begin{figure}[t]
    \centering
    \includegraphics[width=0.48\textwidth]{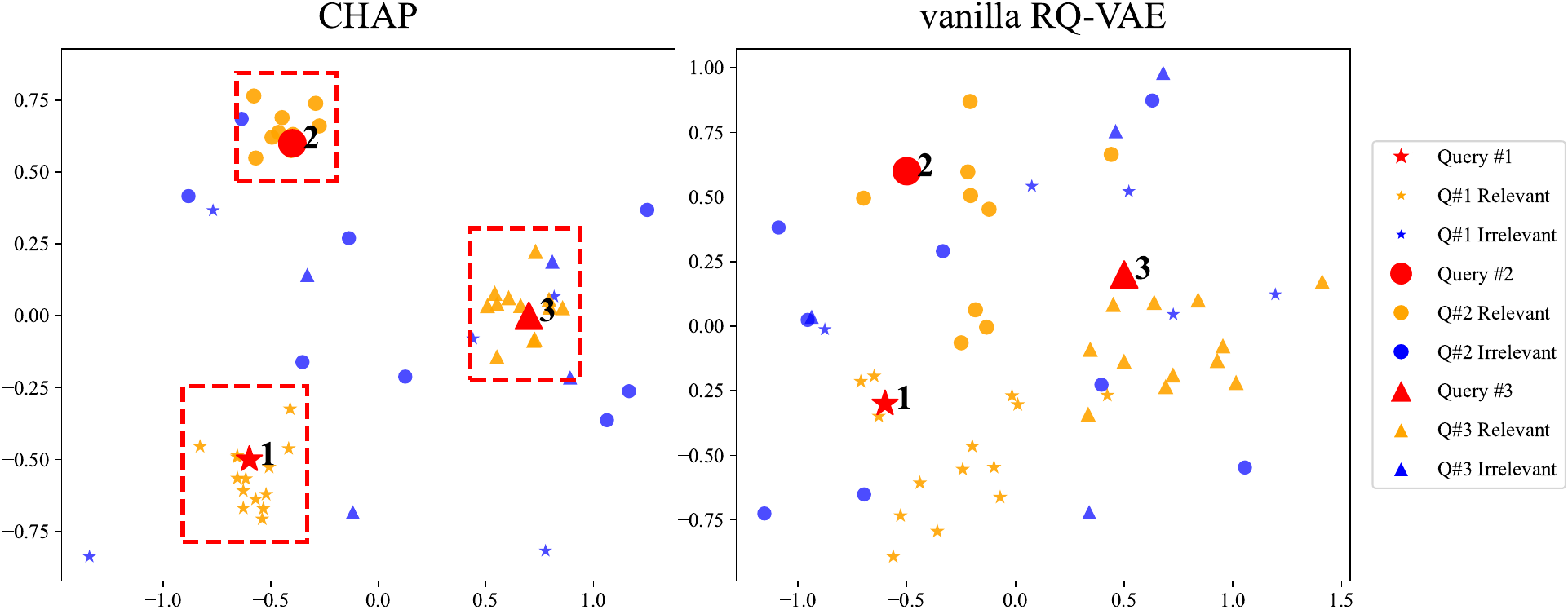}
    \caption{Visualization of Items under three queries.}
    \label{fig:semantic_alignment_comparison}
\end{figure}

Furthermore, visualization of the latent space encoded by DNN encoder of RQ-VAE after PCA dimensionality reduction (from ESCI-us, shown in Fig.~\ref{fig:semantic_alignment_comparison}) intuitively confirms this advantage: while vanilla RQ-VAE suffers from a visible semantic gap with loosely scattered relevant items, CHAP tightly clusters relevant items (orange) directly around their corresponding queries (red), as highlighted by the dashed boxes, while explicitly pushing irrelevant items (blue) further away. This demonstrates the efficacy of our alignment paradigm in drawing dynamic user intents and static item semantics into a highly unified space.
Additional qualitative diagnostics are provided in App.~\ref{app:alignment_diagnostics}.

\begin{figure}[t]
    \centering
    \includegraphics[width=0.95\linewidth]{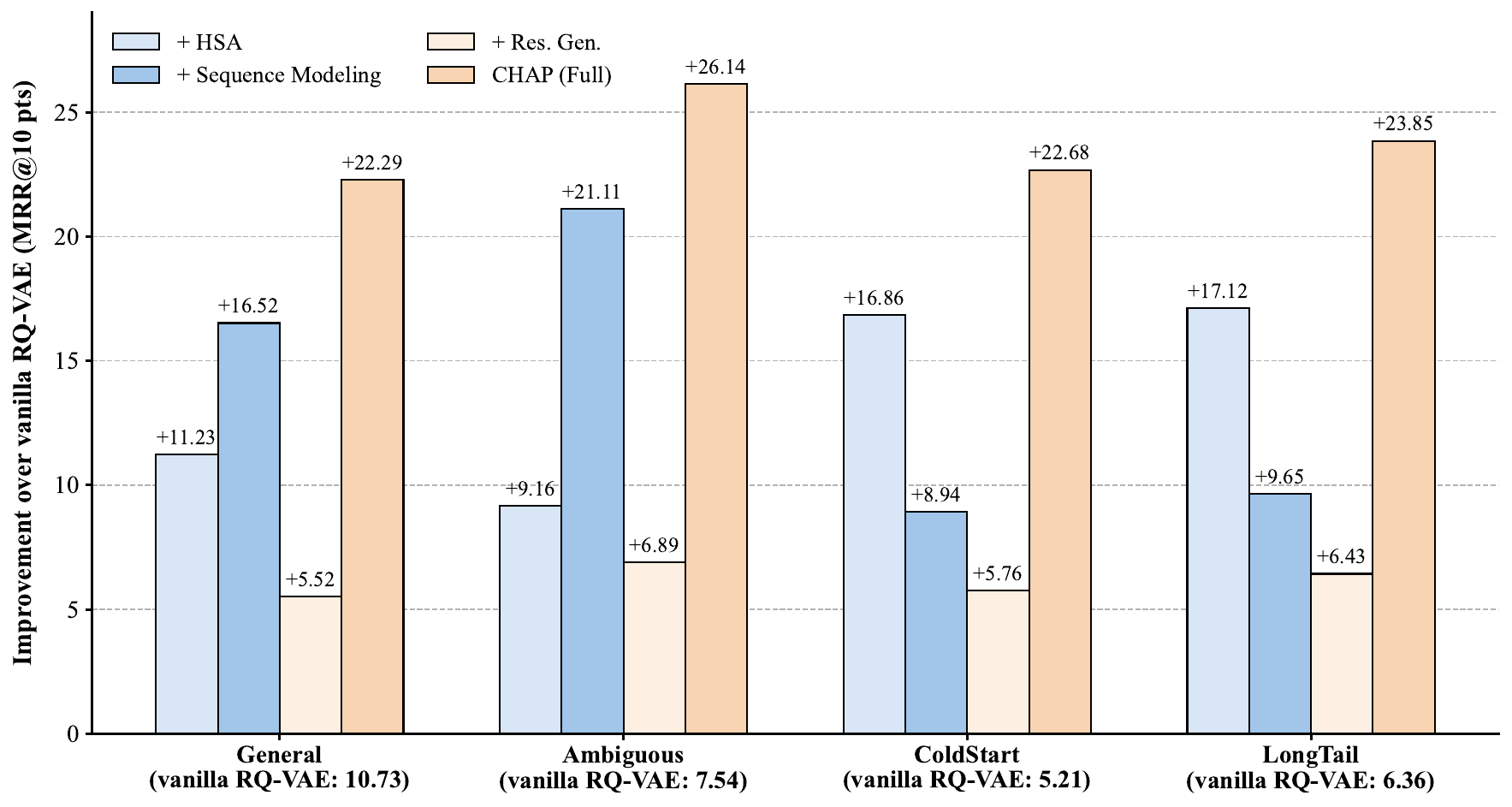}
    \caption{Intent-wise query segment analysis.}
    \label{fig:intent_wise_query_analysis}
\end{figure}

\subsubsection{Intent-wise Query Analysis (RQ3)} 
To further dissect the source of CHAP's superiority, we extract four intent-wise query segments from the Local-Life dataset: General, Ambiguous, ColdStart, and LongTail (detailed in App.~\ref{app:segments}). We explicitly evaluate the absolute MRR@10 improvements brought by individually integrating each core component over the vanilla RQ-VAE baseline (Fig.~\ref{fig:intent_wise_query_analysis}).

The segment-wise results corroborate our design intuitions, revealing the orthogonal strengths of each module. Introducing the dual-view Sequence Modeling delivers the most dominant boost on the \textbf{Ambiguous} segment. This confirms that for highly generic queries, dense historical contextualization is indispensable for intent disambiguation. However, its isolated contribution drops on the \textbf{ColdStart} and \textbf{LongTail} segments, as cold entities inherently lack sufficient interaction histories for sequence models to exploit. Conversely, integrating the HSA module addresses these sparse segments, suggesting that proactive semantic mapping bridges the representation gap for rare queries and new items, independent of popularity biases. Meanwhile, the Residual Generation mechanism yields a stable, structural baseline uplift across all slices by mitigating layer-wise decoding decay.

\input{Sections/Tables/efficiency.tex}

\subsubsection{Efficiency Analysis (RQ4)}
We deeply deconstruct our decoding architecture to evaluate its necessity and efficiency. Degenerating to a standard autoregressive paradigm (w/o Residual Gen.) harms accuracy and halves inference throughput (QPS), indicating our residual design is a powerful decoding accelerator. Removing the decoder's self-attention (w/o Self-Attention) improves QPS with minimal precision loss, showing explicit residual pathways effectively replace self-attention for hierarchical modeling. However, entirely removing the decoder (w/o Decoder) maximizes QPS but collapses MRR. This highlights the indispensable role of cross-attention for dynamic history routing across lengthy user sequences.

To further contextualize the efficiency (Tab.~\ref{tab:efficiency_comparison}), we evaluated the computational overhead using an 8$\times$A100 (80G) GPU cluster for training and a single A100 (80G) GPU for inference QPS. During training, advanced baselines like NCI and LTRGR incur excessive computational overhead due to massive query augmentation or complex list-wise optimization. In contrast, CHAP maintains a highly competitive duration, presenting a cost-effective trade-off for modeling dual-view sequences. During online inference, while standard GR models and dense-heavy frameworks (e.g., COBRA) are severely bottlenecked by repeated layer-wise decoding, our decoupled architecture, combined with parallel sampling, achieves substantially higher QPS, satisfying the strict low-latency requirements of large-scale industrial serving.

\begin{figure}[t]
    \centering
    \begin{subfigure}{0.49\linewidth}
        \centering
        \includegraphics[width=\linewidth, trim=20bp 20bp 22bp 15bp, clip]{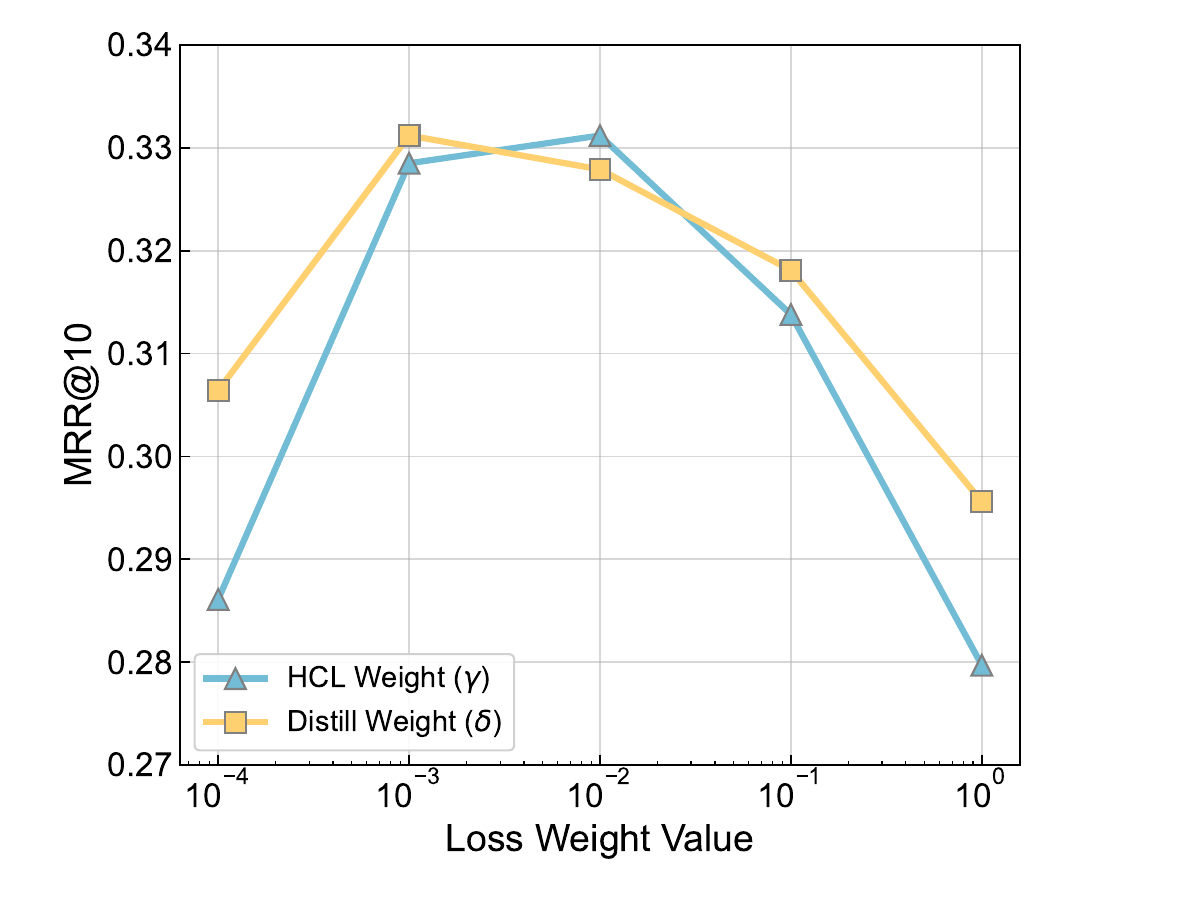}
        \caption{Alignment Weights}
        \label{fig:align_weight}
    \end{subfigure}
    \hfill
    \begin{subfigure}{0.49\linewidth}
        \centering
        \includegraphics[width=\linewidth, trim=20bp 20bp 22bp 15bp, clip]{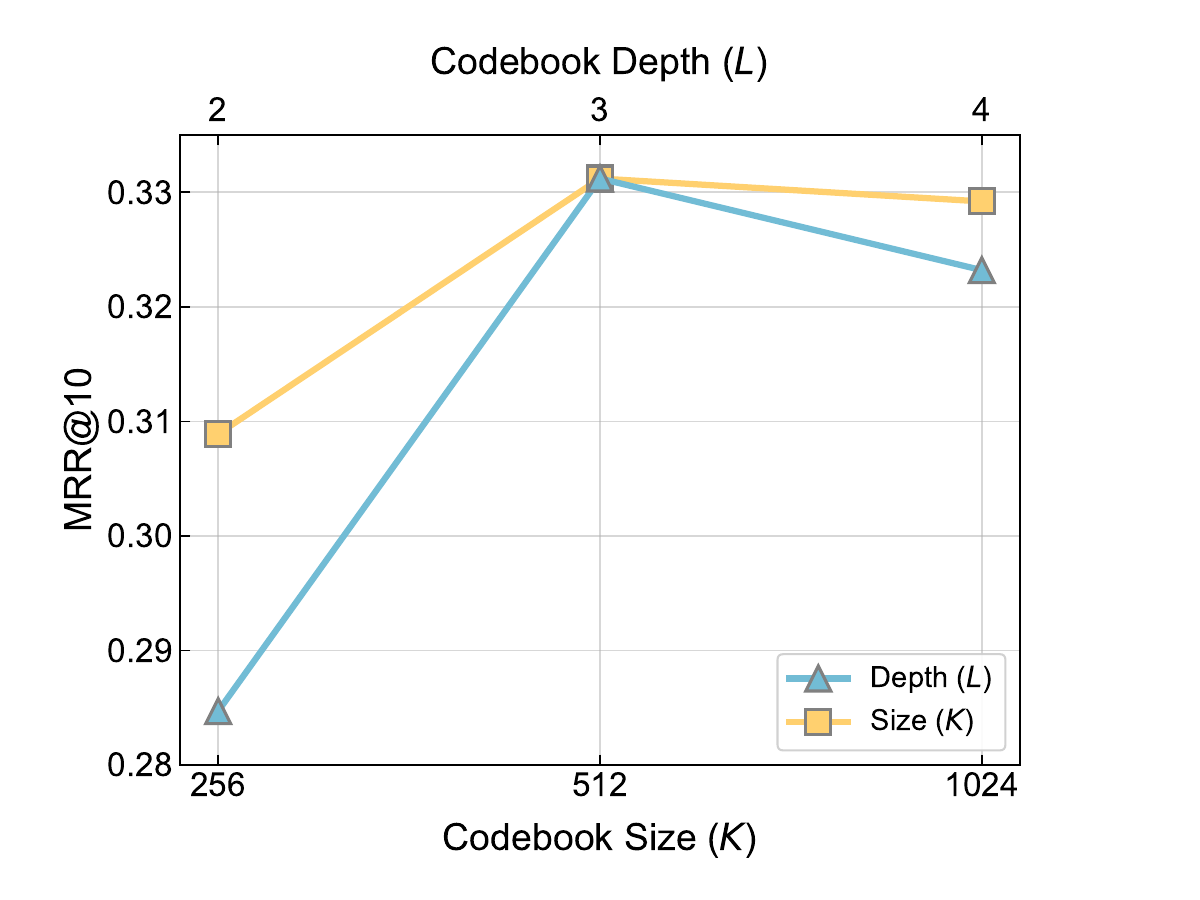}
        \caption{Codebook Size}
        \label{fig:fusion_weight}
    \end{subfigure}
    \caption{Effect of hyperparameters on retrieval tasks.}
    \label{fig:hyper}
\end{figure}

\begin{figure}[t]
    \centering
    \includegraphics[width=0.95\linewidth]{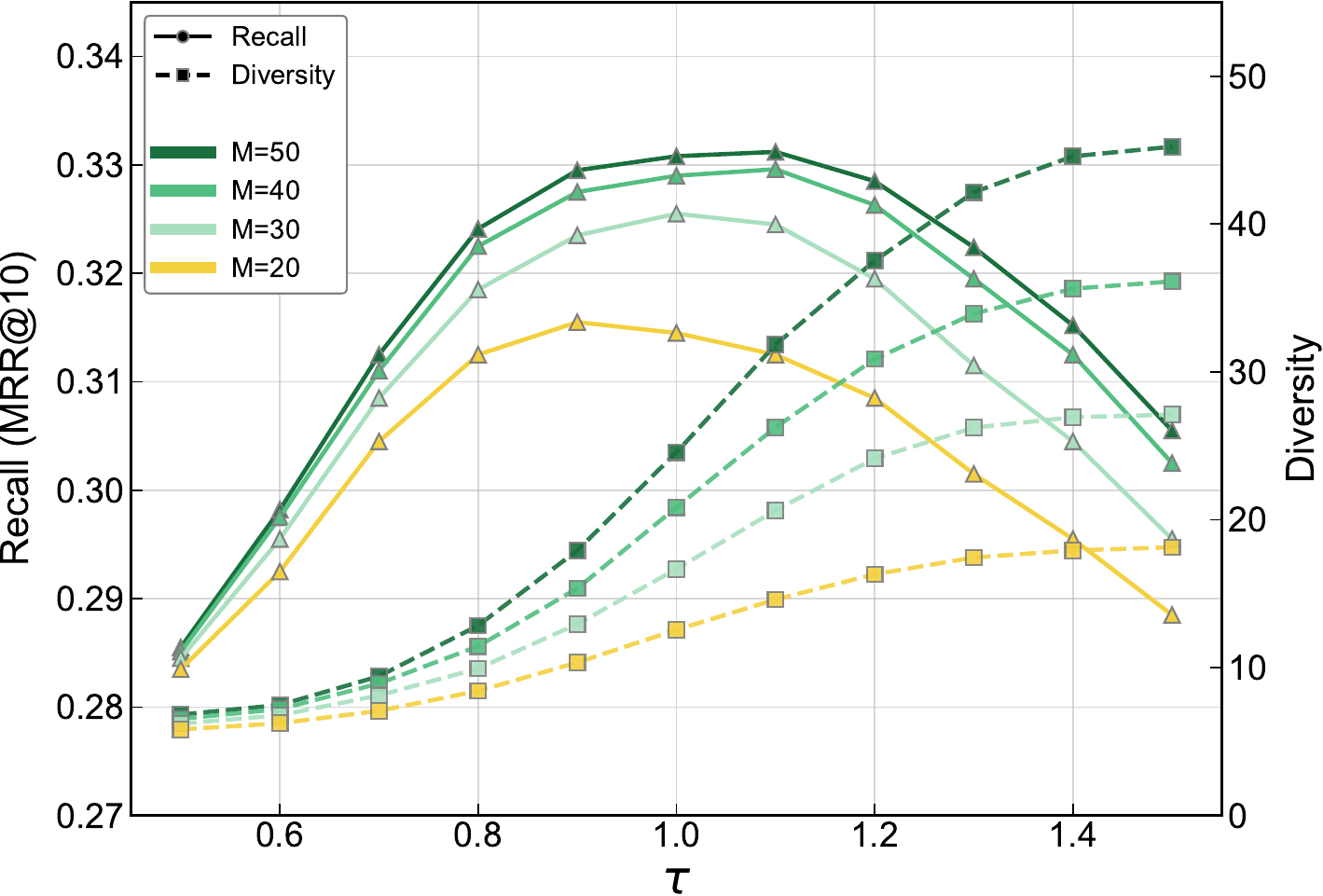}
    \caption{Impact of normalization temperature $\tau$ and candidate set width $M$ on accuracy and diversity.}
    \label{fig:acc_div}
\end{figure}

\subsection{Hyperparameter Analysis (RQ5)}
We analyze the sensitivity of critical hyperparameters on the Local-Life dataset to evaluate CHAP's robustness. Regarding training objectives, the alignment weights exhibit an inverted U-shape trend, peaking at $\gamma=0.01$ and $\delta=0.001$ to achieve optimal semantic alignment without disrupting the primary reconstruction stability. For the codebook scale, performance peaks at depth $L=3$ and size $K=512$. Notably, CHAP demonstrates exceptional robustness to these structural parameters; while extreme settings induce minor metric fluctuations, our framework avoids the performance collapse typically observed in standard GR models under suboptimal codebook assignments. During inference, we regulate the accuracy-diversity equilibrium by varying the normalization temperature $\tau$ across parallel sampling sizes $M$. Ranking accuracy follows an inverted U-shape, suffering from over-confident sharp distributions at low $\tau$ and excessive exploration noise at high $\tau$. Ultimately, CHAP achieves an optimal balance between semantic discrimination and candidate diversity at $\tau=1.1$ with $M=50$. Detailed analysis is provided in Fig.~\ref{fig:hyper} and Fig.~\ref{fig:acc_div}.

\input{Sections/Tables/online.tex}
\subsection{Online A/B Test (RQ6)} 
To verify CHAP's practical effectiveness and scalability, we conducted a rigorous 14-day online A/B test on a leading production local-life service platform, covering 20\% of user traffic. CHAP was deployed via NVIDIA Triton to decouple modules for high-throughput real-time inference, rather than relying on query caching. Compared to the highly optimized online production retrieval system, CHAP achieved consistent improvements: \textbf{UV-CTR} increased by 0.77\%, demonstrating that HSA enhances semantic relevance. 
Moreover, conversion metrics saw substantial lifts, with \textbf{UV-CXR} rising by 2.09\% and \textbf{Pay Order Volume} growing by 2.98\%, showing that CHAP not only attracts user clicks but, crucially, accurately captures latent user intent to drive final purchasing decisions, yielding tangible business growth in industrial scenarios.

%% file: Sections/Tables/dataset_stat.tex
\begin{table}[t]
    \centering
    \caption{Statistics of the datasets in our experiments.}
    \label{tab:dataset_stat}
    \resizebox{\linewidth}{!}{%
    \setlength\tabcolsep{8pt} 
    \begin{tabular}{c|c c c c}
        \toprule
        \textbf{Dataset} & \textbf{\#Items} & \textbf{\#Queries} & \textbf{\#Q-I Pairs} & \textbf{Avg. Hist.} \\
        \midrule
        ESCI-us          & 1,215,854   & 97,346     & 1,818,825  & - \\
        KuaiSearch       & 6,634,118  & 295,561  & 682,228  & 10.37 \\
        Amazon      & 35,772   & 9,070    & 9,070    & 40.12 \\
        Local-Life       & 10,145,542  & 3,089,026 & 3,089,026 & 7.47 \\
        \bottomrule
    \end{tabular}%
    }
\end{table}

%% file: Sections/Tables/overall_result.tex
\begin{table*}[t]
\centering
\caption{Performance comparison on four datasets. \textbf{Bold} indicates the result significantly outperforms all baselines with paired t-tests at $p < 0.05$ level, and \underline{underlined} values denote the second-best results.}
\label{tab:main_results}
\resizebox{\textwidth}{!}{%
\begin{tabular}{c|c|ccc|ccc|ccc|ccc}
\toprule
\multirow{2}{*}{\textbf{Type}} & \multicolumn{1}{c|}{\multirow{2}{*}{\textbf{Model}}} & \multicolumn{3}{c|}{\textbf{ESCI-us}} & \multicolumn{3}{c|}{\textbf{KuaiSearch}} & \multicolumn{3}{c|}{\textbf{Amazon}} & \multicolumn{3}{c}{\textbf{Local-Life}} \\ \cmidrule(lr){3-5} \cmidrule(lr){6-8} \cmidrule(lr){9-11} \cmidrule(lr){12-14}
    & \multicolumn{1}{c|}{} & R@10 & R@50 & NDCG@50 & R@10 & R@50 & HR@50 & R@10 & R@50 & NDCG@50 & R@10 & R@50 & MRR@10 \\ \midrule

\multirow{3}{*}{Sparse} & BM25 & 0.0480 & 0.0932 & 0.0811 & 0.0706 & 0.1564 & 0.2088 & 0.4807 & 0.6780 & 0.3887 & 0.2531 & 0.3415 & 0.0997 \\
    & Doc2Query & 0.0551 & 0.1410 & 0.1015 & 0.0784 & 0.1772 & 0.2382 & 0.5064 & 0.6904 & 0.4039 & 0.2971 & 0.3959 & 0.1374 \\
    & DeepCT & \underline{0.0626} & 0.1486 & 0.0997 & 0.0760 & 0.1898 & 0.2531 & 0.5869 & 0.7622 & 0.4708 & 0.3718 & 0.4387 & 0.1913 \\ \midrule

\multirow{3}{*}{Dense} & DPR & 0.0552 & 0.1765 & 0.1104 & 0.0826 & 0.2079 & 0.2769 & 0.2010 & 0.3413 & 0.1666 & 0.3153 & 0.4820 & 0.1582 \\
    & Sen-T5 & 0.0458 & 0.1363 & 0.0845 & 0.0673 & 0.1622 & 0.2227 & 0.5690 & 0.7967 & 0.4376 & 0.2360 & 0.3802 & 0.0956 \\
    & MPNet & 0.0294 & 0.0879 & 0.0531 & 0.0619 & 0.1279 & 0.1970 & 0.5961 & 0.7755 & 0.4719 & 0.1820 & 0.3210 & 0.0815 \\ \midrule

\multirow{2}{*}{Personalized} & TEM & 0.0213 & 0.0545 & 0.0402 & 0.0852 & 0.2131 & 0.2776 & 0.4814 & 0.7301 & 0.3535 & 0.3551 & 0.5401 & 0.1716 \\
    & CoPPS & 0.0246 & 0.0878 & 0.0458 & 0.0877 & 0.2153 & \underline{0.2832} & 0.4854 & 0.8004 & 0.3699 & 0.4265 & 0.5971 & 0.2403 \\ \midrule

\multirow{7}{*}{Generative} & DSI & 0.0217 & 0.0552 & 0.0433 & 0.0623 & 0.1369 & 0.2018 & 0.4181 & 0.6202 & 0.3414 & 0.1868 & 0.3056 & 0.0836 \\
    & NCI & 0.0432 & 0.1006 & 0.0803 & 0.0591 & 0.1270 & 0.1781 & 0.3765 & 0.5836 & 0.3139 & 0.2540 & 0.3951 & 0.1014 \\
    & LTRGR & 0.0316 & 0.0944 & 0.0629 & 0.0688 & 0.1501 & 0.2184 & 0.5143 & 0.7025 & 0.4062 & 0.2608 & 0.4644 & 0.1120 \\
    & TIGER & 0.0487 & 0.1335 & 0.0814 & 0.0731 & 0.1796 & 0.2427 & 0.5387 & 0.7292 & 0.4237 & 0.2659 & 0.4068 & 0.1073 \\
    & MERGE & 0.0612 & 0.1903 & 0.1070 & 0.0858 & 0.2033 & 0.2625 & 0.5697 & 0.7819 & 0.4523 & 0.3850 & 0.5828 & 0.2389 \\
    & COBRA & 0.0585 & \underline{0.1950} & \underline{0.1126} & \underline{0.0890} & \underline{0.2074} & 0.2803 & \underline{0.6205} & \underline{0.8474} & \underline{0.4906} & \underline{0.5020} & \underline{0.6057} & \underline{0.2810} \\ \cmidrule(l){2-14}
    & \textbf{CHAP (Ours)} & \textbf{0.1127} & \textbf{0.2452} & \textbf{0.2454} & \textbf{0.0944} & \textbf{0.2289} & \textbf{0.2949} & \textbf{0.7358} & \textbf{0.9416} & \textbf{0.5191} & \textbf{0.5803} & \textbf{0.8085} & \textbf{0.3302} \\ \bottomrule
\end{tabular}%
}
\end{table*}

%% file: Sections/Tables/ablation.tex
\begin{table}[t]
    \centering
    \caption{Ablation study of CHAP on Local-Life.}
    \label{tab:ablation}
    \resizebox{\linewidth}{!}{%
    \setlength{\tabcolsep}{6pt}
    \begin{tabular}{l|cccc}
    \toprule
    \textbf{Variant} & R@10 & R@50 & MRR@10 & QPS \\ \midrule
    \textbf{CHAP (Full Model)} & \textbf{0.5803} & \textbf{0.8089} & \textbf{0.3302} & 78.9 \\ \midrule
    \multicolumn{5}{l}{\textit{(A) Hierarchical Semantic Alignment (HSA)}} \\
    \ w/o HSA & 0.5115 & 0.7488 & 0.2845 & 78.9 \\
    \quad w/o $\mathcal{L}_\mathrm{CSA}$ & 0.5164 & 0.7602 & 0.2721 & 78.9 \\
    \quad w/o $\mathcal{L}_\mathrm{HCL}$ & 0.5218 & 0.7534 & 0.2858 & 78.9 \\
    \quad w/o $\mathcal{L}_\mathrm{Distill}$ & 0.5692 & 0.7895 & 0.3063 & 78.9 \\ \midrule
    \multicolumn{5}{l}{\textit{(B) Sequence Modeling}} \\
    \ w/o Dense Refinement & 0.4352 & 0.6455 & 0.2241 & 81.2 \\ \midrule
    \multicolumn{5}{l}{\textit{(C) Residual Cascading Generation}} \\
    \ w/o Residual Gen. \textit{(Autoregressive)} & 0.5285 & 0.7610 & 0.2849 & 40.5 \\
    \ w/o Self-Attention \textit{(Cross-Attention)} & 0.5712 & 0.8062 & 0.3237 & 84.8 \\
    \ w/o Decoder \textit{(Pooling + MLP)} & 0.4651 & 0.6836 & 0.2584 & 112.4 \\ 
    \bottomrule
    \end{tabular}%
    }
\end{table}

%% file: Sections/Tables/efficiency.tex
\begin{table}[t]
    \centering
    \caption{Efficiency comparison on Local-Life. ItemID and Seq. denote the training time for ItemID construction and the GR model, respectively. Total is their sum.}
    \label{tab:efficiency_comparison}
    \resizebox{\linewidth}{!}{%
    \setlength{\tabcolsep}{4pt}
    \begin{tabular}{c|ccc|c}
    \toprule
    \textbf{Model} & \textbf{ItemID (H)} & \textbf{Seq. (H)} & \textbf{Total (H)} & \textbf{QPS} \\ \midrule
    DSI & 1.3 & 16.5 & 17.8 & 37.5 \\
    NCI & 1.3 & 34.0 & 35.3 & 26.2 \\
    LTRGR & 1.5 & 31.9 & 33.4 & 34.7 \\
    TIGER & 0.4 & 18.2 & 18.6 & 39.2 \\
    MERGE & 0.8 & 19.3 & 20.1 & 36.8 \\
    COBRA & 0.4 & 26.4 & 26.8 & 32.4 \\ \midrule
    \textbf{CHAP (Ours)} & 0.5 & 23.8 & 24.3 & 78.9 \\ 
    \bottomrule
    \end{tabular}%
    }
    \end{table}

%% file: Sections/Tables/online.tex
\begin{table}[t]
\centering
\caption{Online A/B testing results over a 14-day period. All metrics are statistically significant with paired t-tests at $p < 0.05$ level.}
\label{tab:online}
\resizebox{0.8\linewidth}{!}{%
\begin{tabular}{c c}
\toprule
\textbf{Metric} & \textbf{Relative Improvement} \\ \midrule
UV-CTR (Click-Through Rate) & +0.77\% \\
UV-CXR (Conversion Rate) & +2.09\% \\
Pay Orders (Order Volume) & +2.98\% \\ \bottomrule
\end{tabular}%
}
\end{table}

%% file: Sections/6_Conclusion.tex
In this work, we propose CHAP, a novel personalized GR framework designed to overcome the inherent semantic gap and inference latency bottlenecks of existing GR paradigms. By introducing the HSA module, CHAP effectively maps dynamic user intents directly into the static item quantization space. Furthermore, our dual-view sequence modeling, coupled with a single-pass Residual Cascading Generation mechanism, decouples historical intent routing from layer-wise decoding. This architecture not only ensures fine-grained semantic precision but also acts as a powerful decoding accelerator. Extensive experiments across four diverse real-world datasets, alongside a rigorous online A/B test on a leading local-life platform, demonstrate that CHAP establishes a new state-of-the-art. It achieves a balance between exact intent alignment and industrial-grade inference throughput.

%% file: Sections/7_Limitations.tex
Although CHAP demonstrates strong empirical effectiveness and industrial efficiency, several limitations remain for future exploration. First, in the HSA stage, freezing the item-side codebook provides a stable quantization manifold and prevents semantic drift. However, this design may also prevent item representations from being dynamically updated with collaborative filtering signals or query-side feedback, which could limit the upper bound of personalization in highly dynamic environments.
Second, our dual-view personalized sequence modeling is mainly validated with hierarchical SIDs generated by RQ-VAE-style quantization. While this setting is representative for recent generative retrieval systems, the generalization ability of the overall framework to other types of ItemIDs, such as lexical identifiers, learned atomic IDs, or tree-based semantic IDs, still requires systematic investigation.
Finally, the Residual Cascading Generation module substantially improves inference efficiency by avoiding repeated heavy decoder passes. Nevertheless, how the acceleration ratio evolves when scaling model size, codebook depth, and candidate budget remains an open question. Further compression techniques, such as model distillation for the residual generation blocks, may provide an additional path toward more efficient large-scale deployment.

%% file: Sections/8_Ethical_Considerations.tex
This research was reviewed and approved by the relevant technical ethics committee. All data collection procedures followed the platform's compliance requirements and were conducted with user permission or authorization where applicable. Before being used for model training, evaluation, and analysis, all user-related data were anonymized or de-identified to remove personally identifiable information. Our experiments rely only on aggregated statistics and anonymized behavioral signals, and do not attempt to infer or recover individual identities. The online A/B test was conducted under the platform's standard safety and compliance protocols to minimize potential risks to user experience. The open-source implementation of CHAP is released under a CC BY-NC-SA 4.0 license, explicitly disclaiming liability for direct commercial deployment to ensure strict non-commercial, research-only usage.

%% file: Sections/Appendix.tex
\section{Related Works}
\label{app:related_works}

\textbf{Sparse retrieval} methods like BM25~\cite{robertson2009probabilistic} and TF-IDF~\cite{ramos2003using} rely on exact term matching but suffer from vocabulary mismatch. Conversely, \textbf{dense retrieval} approaches (e.g., DPR~\cite{karpukhin2020dense}, ANCE~\cite{xiong2020approximate}) map inputs into a shared vector space, enhancing semantic matching but necessitating costly external indices like HNSW~\cite{malkov2018efficient}. To address these limitations, STAR~\cite{zhan2021optimizing} introduces stable negative sampling, while ColBERT~\cite{khattab2020colbert} and Sentence-BERT~\cite{reimers2019sentence} refine fine-grained similarity and semantic understanding. Recent work also investigates scaling laws in dense models~\cite{fang2024scaling}.

\textbf{Personalized retrieval} tailors results by incorporating user contexts. QEM~\cite{ai2019zero} investigates personalization mechanisms, while DREM~\cite{ai2019explainable} models dynamic user-product relationships via knowledge graphs. Embedding-based approaches like ZAM~\cite{ai2019zero}, HEM~\cite{ai2017learning}, and TEM~\cite{bi2020transformer} integrate user profiles and history into unified embeddings. In conversational domains, CHIQ~\cite{mo2024chiq} enhances accuracy by leveraging historical search queries. TTP~\cite{jiang2026think} introduces a unified reasoning and retrieval framework for personalized dense retrieval.

\textbf{Generative retrieval} internalizes corpus knowledge to directly generate item identifiers. DSI~\cite{tay2022transformer} pioneered this paradigm, followed by semantic enhancements in SE-DSI~\cite{tang2023semantic}. Various optimization strategies have emerged: SEATER~\cite{si2024generative} utilizes balanced k-ary trees; Tied-Atomic~\cite{nguyen2023generative} connects GR with dense retrieval; GenRRL~\cite{zhou2023enhancing} employs reinforcement learning; and LTRGR~\cite{li2024learning} incorporates ranking tasks. TIGER~\cite{rajput2023recommender} first utilized RQ-VAE for SIDs. Regarding architecture, COBRA~\cite{yang2025sparse} integrates sparse and dense vectors, while Hi-gen~\cite{wu2024hi}, CAT-ID$^2$~\cite{liu2025cat}, and MERGE~\cite{zhang2025multi} optimize ID generation by exploiting multi-level correlations.

\textbf{Recent and concurrent studies} have further revisited Semantic ID based generative retrieval from several complementary perspectives. A line of work improves the construction of Semantic IDs beyond item-only reconstruction. For example, CQ-SID introduces category-aware and query-item contrastive objectives for e-commerce generative retrieval~\cite{zhu2026efficient}, while DIGER~\cite{fu2026differentiable} and UniSID~\cite{jiang2026end} explore differentiable or end-to-end Semantic ID learning to better align identifier assignment with downstream recommendation objectives. Another direction studies the dynamics and scalability of Semantic IDs, such as mitigating collaborative Semantic ID staleness under evolving user-item interactions~\cite{baikalov2026mitigating} and exploiting longer Semantic IDs to enhance representation capacity while controlling decoding cost~\cite{xia2026unleash}. Closely related to our efficiency motivation, MLP-based distilled generative recommenders show that attention-heavy autoregressive decoders can be distilled into lightweight MLP-based heads for faster Semantic ID generation~\cite{guo2026mlps}, suggesting that much of the decoder-side computation is redundant once global user context and prefix structure are properly preserved. These studies are largely orthogonal to CHAP: they mainly focus on identifier learning, temporal ID maintenance, or decoder distillation, whereas CHAP targets personalized generative retrieval by explicitly aligning query representations with the item quantization path, jointly modeling sparse Semantic IDs and dense representations, and using residual cascading generation to reduce autoregressive latency while preserving fine-grained personalized relevance.

\section{Proof of Cascaded Sparse-Dense Modeling}
\label{app:cascade_proof}

\textbf{Theorem 1 (Entropy reduction of CHAP's cascaded modeling).}
Let $X$ denote the personalized context encoded from $X_{in}$, and let
$\mathbf{C}_i=(C_i^0,\ldots,C_i^{L-1})$ and $\mathbf{V}_i$ denote the random
variables of the target item's hierarchical SID and dense vector, respectively.
Consider an independent dual-view factorization
\begin{align}
    P_{\mathrm{ind}}(\mathbf{C}_i,\mathbf{V}_i|X)
    &=
    \prod_{l=0}^{L-1} P(C_i^l|X)
    \cdot P(\mathbf{V}_i|X),
\end{align}
and CHAP's cascaded sparse-dense factorization
\begin{align}
    P_{\mathrm{chap}}(\mathbf{C}_i,\mathbf{V}_i|X)
    &=
    \prod_{l=0}^{L-1} P(C_i^l|C_i^{<l},X)
    \notag \\
    &\quad \cdot P(\mathbf{V}_i|\mathbf{C}_i,X).
\end{align}
Denote their total uncertainties by $\mathcal{H}_{\mathrm{ind}}$ and
$\mathcal{H}_{\mathrm{chap}}$, where $H(\cdot)$ is used for discrete SIDs and
$h(\cdot)$ for the continuous dense vector. Then
\begin{equation}
    \mathcal{H}_{\mathrm{chap}}(\mathbf{C}_i,\mathbf{V}_i|X)
    \le
    \mathcal{H}_{\mathrm{ind}}(\mathbf{C}_i,\mathbf{V}_i|X).
\end{equation}
Equality holds if and only if each SID level is conditionally independent of
its prefix given $X$, and $\mathbf{V}_i$ is conditionally independent of
$\mathbf{C}_i$ given $X$.

\textit{Proof.}
Writing $P_{\mathrm{ind}}$ and $P_{\mathrm{chap}}$ for the two distributions
above, the independent dual-view entropy decomposes as
\begin{align}
    \mathcal{H}_{\mathrm{ind}}
    &= -\mathbb{E}_{P_{\mathrm{ind}}}[\log P_{\mathrm{ind}}]
    \notag \\
    &= \sum_{l=0}^{L-1} H(C_i^l|X)
    \notag \\
    &\quad + h(\mathbf{V}_i|X).
\end{align}
For CHAP's cascaded factorization, we have
\begin{align}
    \mathcal{H}_{\mathrm{chap}}
    &= -\mathbb{E}_{P_{\mathrm{chap}}}[\log P_{\mathrm{chap}}]
    \notag \\
    &= \sum_{l=0}^{L-1} H(C_i^l|C_i^{<l},X)
    \notag \\
    &\quad + h(\mathbf{V}_i|\mathbf{C}_i,X).
\end{align}
By the non-negativity of conditional mutual information,
\begin{equation}
    H(C_i^l|X) - H(C_i^l|C_i^{<l},X)
    = I(C_i^l; C_i^{<l}|X) \ge 0,
\end{equation}
and
\begin{equation}
    h(\mathbf{V}_i|X) - h(\mathbf{V}_i|\mathbf{C}_i,X)
    = I(\mathbf{V}_i;\mathbf{C}_i|X) \ge 0.
\end{equation}
Substituting these inequalities into the two entropy decompositions yields
\begin{align}
    \mathcal{H}_{\mathrm{ind}} - \mathcal{H}_{\mathrm{chap}}
    &=
    \sum_{l=0}^{L-1} I(C_i^l; C_i^{<l}|X)
    \notag \\
    &\quad + I(\mathbf{V}_i;\mathbf{C}_i|X) \ge 0.
\end{align}
Therefore, CHAP's coarse-to-fine SID generation followed by dense refinement
does not increase the modeling uncertainty compared with independently
predicting sparse and dense views. The equality condition follows directly from
the equality condition of conditional mutual information: all terms above are
zero only when $C_i^l \perp C_i^{<l}\,|\,X$ for every level $l$ and
$\mathbf{V}_i \perp \mathbf{C}_i\,|\,X$. This completes the proof.

\section{Dataset Details}
\label{app:dataset_details}

In this section, we provide detailed descriptions of the four datasets used to evaluate the proposed CHAP framework. The detailed statistics are summarized in Tab.~\ref{tab:open_source_dataset_stat_appendix}.

\input{Sections/Tables/open_source_dataset_stat_appendix.tex}

\textbf{ESCI-us.} We utilize the large version of the English subset from the ESCI\footnote{\url{https://github.com/amazon-science/esci-data}}~\cite{reddy2022shopping} dataset. Unlike the widely used Amazon Product Reviews~\cite{he2016ups} dataset, which necessitates constructing synthetic queries, ESCI contains authentic queries derived from actual user logs on Amazon, filled with the noise and complexity of real-world search traffic. Each query in ESCI is associated with products labeled by relevance levels: Exact (E), Substitute (S), Complement (C), and Irrelevant (I). This dataset is characterized by complex linguistic patterns, including negation conditions (e.g., ``not'' and ``without'') and intricate attribute constraints (e.g., dimensions, price, functionality). These factors result in a low text overlap between queries and product descriptions, making it a rigorous benchmark for evaluating a model's semantic understanding capabilities. During sequence modeling, we treat items labeled ``E'' and ``S'' as positive samples. Notably, unlike existing works~\cite{liu2025cat,zhang2025multi} that utilized the small version of the dataset and applied specific cleaning and preprocessing steps, we directly adopt the large version with raw query and item texts. We strictly follow the official pre-defined train and test splits without any additional preprocessing to rigorously test the model's robustness and generalization capabilities.

\textbf{KuaiSearch.} KuaiSearch\footnote{\url{https://github.com/benchen4395/KuaiSearch}}~\cite{li2026kuaisearch} is a large-scale e-commerce search dataset built upon real user search interactions from the Kuaishou platform. In contrast to existing datasets that often rely on heuristically constructed queries or aggressively filter out inactive entities, KuaiSearch preserves authentic user queries and natural-language product texts without popularity-based filtering. This ensures the retention of cold-start users and long-tail products, accurately reflecting real-world e-commerce long-tail distributions. The whole dataset contains approximately 330,000 users, 18,000,000 products, and 2,500,000 real search queries. Furthermore, it provides extensive long-term user behavior sequences and systematically covers the core stages of industrial search pipelines. In our experiments, we specifically utilize the \textit{recall} subset. During data construction, we treat both clicked and purchased items as positive samples. Strictly following the methodology described in the original paper, we apply identical data cleaning steps—filtering out entries that lacked explicit search queries or had no positive samples—and subsequently partition the cleaned dataset into training and testing sets using a random 9:1 split.

\textbf{Amazon.} To evaluate our model's capacity for fine-grained personalized retrieval, we utilize the Amazon search dataset from the Personalized Web Agent Benchmark (PersonalWAB)\footnote{\url{https://github.com/HongruCai/PersonalWAB}}~\cite{cai2025large}. Unlike traditional static search benchmarks, PersonalWAB is explicitly designed to assess the integration of user profiles and historical web behaviors in understanding customized instructions. The dataset originates from real-world Amazon interaction logs and encompasses diverse user profiles alongside tens of thousands of historical web behaviors. It features highly personalized, human-synthesized natural language instructions tailored to each user's profile, requiring models to dynamically infer specific user preferences. This dataset presents a unique challenge for generative retrieval models, as it necessitates a deep semantic fusion of the current query, user profiles, and heterogeneous historical behaviors to achieve accurate personalized product search. To fully preserve the complexity and noise of these customized instructions, we utilized this dataset in its original form and directly adopted the officially pre-defined train and test splits without performing any data cleaning or preprocessing.

\textbf{Local-Life.} To assess performance in a dynamic, personalized environment, we introduce a large-scale industrial dataset collected from 30 days of user interaction logs on a leading Local-Life service platform. Compared to public datasets like KuaiSAR~\cite{sun2023kuaisar} or JDSearch~\cite{liu2023jdsearch}, which only provide encrypted token IDs preventing semantic verification, this dataset comprises raw natural language text. This transparency is crucial for verifying our model's ability to align semantics between queries and items. The dataset covers diverse scenarios (e.g., Dining, Entertainment, Travel) and heterogeneous search intents (e.g., Brand Search, Ambiguous Intent, Exact Item Search, and Long-tail or Cold-start Search), comprehensively reflecting real-world user behaviors. We construct interaction sequences from the first 29 days for training and the last day for testing.

\section{Baseline Descriptions}
\label{app:baselines}

To rigorously evaluate the proposed CHAP framework, we comprehensively compare it against 14 representative baselines categorized into sparse retrieval, dense retrieval, personalized retrieval, and generative retrieval (GR) methods. 

\textbf{Sparse Retrieval Methods.} 
\begin{itemize}
    \item \textbf{BM25}~\cite{robertson2009probabilistic}: A classical probabilistic information retrieval model that relies on exact lexical matching based on term frequency and inverse document frequency.
    \item \textbf{Doc2Query}~\cite{nogueira2019document}: A document expansion technique that utilizes a sequence-to-sequence model to predict potential queries for each document prior to indexing, thereby enriching the vocabulary and mitigating the vocabulary mismatch problem.
    \item \textbf{DeepCT}~\cite{dai2019context}: A deep contextualized term weighting framework that leverages BERT-based text representations to estimate the document-specific semantic importance of terms, translating them into enhanced term weights for first-stage sparse retrieval.
\end{itemize}

\textbf{Dense Retrieval Methods.} 
\begin{itemize}
    \item \textbf{DPR}~\cite{karpukhin2020dense}: Dense Passage Retriever, a highly effective bi-encoder architecture optimized via in-batch and hard negative contrastive learning to map queries and items into a shared semantic vector space.
    \item \textbf{Sen-T5}~\cite{ni2022sentence}: A scalable sentence encoding model built upon the T5 framework, pre-trained on massive text-to-text and contrastive tasks to generate high-quality, continuous text representations.
    \item \textbf{MPNet}~\cite{song2020mpnet}: A robust unified pre-training framework that successfully combines the advantages of masked language modeling (e.g., BERT) and permuted language modeling (e.g., XLNet) for enhanced document embeddings.
\end{itemize}

\textbf{Personalized Retrieval Methods.} 
\begin{itemize}
    \item \textbf{TEM}~\cite{bi2020transformer}: Transformer-based Embedding Model, which explicitly encodes user search history alongside current queries to construct dynamic user profiles, aligning retrieval representations with personalized intents.
    \item \textbf{CoPPS}~\cite{dai2023contrastive}: Contrastive learning for Personalized Product Search, a framework that discriminates fine-grained user intents by applying multi-view contrastive learning over user behavior sequence representations.
\end{itemize}

\textbf{Generative Retrieval (GR) Methods.} 
\begin{itemize}
    \item \textbf{DSI}~\cite{tay2022transformer}: A pioneering end-to-end differentiable search index that directly generates relevant document identifiers using a Transformer memory based on semantic clustering.
    \item \textbf{NCI}~\cite{wang2022neural}: Neural Corpus Indexer, a strong GR baseline that utilizes a prefix-aware decoder and a query generation strategy to optimize hierarchical semantic identifiers effectively.
    \item \textbf{TIGER}~\cite{rajput2023recommender}: A generative recommender and retrieval model that leverages a Residual Quantized Variational AutoEncoder (RQ-VAE) to generate hierarchical, semantically rich discrete IDs strictly based on item textual features.
    \item \textbf{LTRGR}~\cite{li2024learning}: Learning-to-Rank in Generative Retrieval, which integrates ranking-aware principles into the generation process by optimizing the log-probabilities of generation according to target ranking metrics like ListMLE.
    \item \textbf{MERGE}~\cite{zhang2025multi}: A recent generative retrieval framework that aims to optimize identifier assignments by aligning multiple positive target items across varying relevance signals for multi-level ranking.
    \item \textbf{COBRA}~\cite{yang2025sparse}: A state-of-the-art cascading generative framework that unites sparse identifier generation and dense representation regression within a joint optimization landscape to improve retrieval coverage and accuracy.
\end{itemize}

\section{Evaluation Metrics Details}
\label{app:metrics}

To comprehensively evaluate the retrieval accuracy and ranking quality of the proposed CHAP framework, we employ four widely adopted metrics: Recall, Hit Ratio (HR), Mean Reciprocal Rank (MRR), and Normalized Discounted Cumulative Gain (NDCG). Here, we detail their formulations and the specific rationales for selecting them based on the characteristics of different datasets.

\textbf{Recall (R@K).} Recall evaluates the model's ability to retrieve all relevant items. It provides a macroscopic view of coverage, which is essential for measuring overall semantic alignment.
\begin{equation}
    \text{R@}K = \frac{1}{|Q|} \sum_{q \in Q} \frac{|\mathcal{R}_q \cap \hat{\mathcal{I}}_{q,K}|}{|\mathcal{R}_q|},
\end{equation}
where $Q$ is the set of queries, $\mathcal{R}_q$ denotes the ground-truth relevant items for query $q$, and $\hat{\mathcal{I}}_{q,K}$ represents the top-$K$ retrieved candidates.

\textbf{Hit Ratio (HR@K).} Hit Ratio measures the probability that a user's target item successfully appears in the top-$K$ retrieved list. It is formulated as an indicator function:
\begin{equation}
    \text{HR@}K = \frac{1}{|Q|} \sum_{q \in Q} \mathbb{I}(|\mathcal{R}_q \cap \hat{\mathcal{I}}_{q,K}| > 0),
\end{equation}
where $\mathbb{I}(\cdot)$ evaluates to 1 if the condition is true and 0 otherwise. \textit{Rationale:} In practical industrial scenarios (e.g., e-commerce purchases or local-life orders), user intent is often highly focused on a single specific item. HR effectively reflects the success rate of fulfilling this distinct user intent, making it highly suitable for evaluating performance on datasets like Local-Life and KuaiSearch.

\textbf{Mean Reciprocal Rank (MRR@K).} MRR strictly assesses the ranking quality by focusing on the position of the \textit{first} relevant item retrieved:
\begin{equation}
    \text{MRR@}K = \frac{1}{|Q|} \sum_{q \in Q} \frac{1}{\text{rank}_q^*},
\end{equation}
where $\text{rank}_q^*$ is the rank position of the first relevant item in the top-$K$ list (if no relevant item is in the top $K$, the score is 0). \textit{Rationale:} For single-target or exact-intent datasets (such as PersonalWAB and Local-Life), users typically stop browsing once they find the exact match. MRR rigorously penalizes models that place the target item lower down the list, directly correlating with user satisfaction and online CTR.

\textbf{Normalized Discounted Cumulative Gain (NDCG@K).} NDCG evaluates the ranking quality by considering both the presence and the position of multiple relevant items, factoring in multi-level relevance scores:
\begin{equation}
    \text{DCG@}K = \sum_{i=1}^{K} \frac{2^{rel_i} - 1}{\log_2(i+1)},
\end{equation}
\begin{equation}
    \text{NDCG@}K = \frac{\text{DCG@}K}{\text{IDCG@}K},
\end{equation}
where $rel_i$ is the graded relevance score of the item at rank $i$, and IDCG@$K$ is the ideal DCG obtained by perfectly sorting all relevant items. \textit{Rationale:} For datasets like ESCI-us, a single query often maps to multiple items with varying relevance levels (e.g., Exact, Substitute, Complement). NDCG is uniquely capable of capturing these fine-grained grading differences, rewarding models that place highly relevant items at the top while tolerating lower-ranked partially relevant items.

\section{Implementation Details for CHAP and Baselines}
\label{app:implementation}

To ensure reproducibility and address concerns regarding experimental fairness, particularly about candidate generation budgets and ranking protocols, we provide the comprehensive implementation and optimization details for our proposed CHAP framework and all evaluated baseline models.

\subsection{Training Details of CHAP}
The training of CHAP consists of two primary stages. 
In the SID Learning Stage, we pre-train the RQ-VAE to discretize item embeddings. We use the AdamW optimizer with a learning rate of $1 \times 10^{-4}$ and a batch size of 1024, training the autoencoder for 500 epochs with the commitment loss weight $\beta=0.5$ to ensure semantic stability and avoid codebook collapse.
In the Hierarchical Semantic Alignment and Generation Stage, we freeze the pre-trained item codebook and fine-tune the Transformer backbone along with the alignment modules. This stage is trained for 50 epochs using the AdamW optimizer with a reduced learning rate of $1 \times 10^{-5}$, a warm-up ratio of 0.1, and a linear learning rate decay schedule. The maximum sequence length for the dual-view user behavior input is truncated to 512 tokens to balance context modeling and computational memory. All experiments are implemented in PyTorch and conducted on eight NVIDIA A100 (80GB) GPUs.

\subsection{Implementation Settings of Baselines}
To ensure a strictly fair, apples-to-apples comparison, all deep learning-based baselines are equipped with backbones of comparable capacity to CHAP and are provided with identical raw textual features and historical sequence lengths. Specifically, for English datasets, we uniformly use BERT-base and T5-base; for Chinese datasets, we replace them with Chinese-BERT-base and mT5-base to handle the specific tokenization and vocabulary. BERT-base has about 110M parameters, Chinese-BERT-base about 102M, T5-base about 220M, and mT5-base about 580M. All GR baselines reported in Tab.~\ref{tab:efficiency_comparison} use the same-capacity T5-base or mT5-base backbone as CHAP, so the efficiency comparison mainly reflects framework-level design differences rather than backbone scale.

\textbf{Sparse Retrieval.} We utilize the \texttt{Pyserini} toolkit for BM25, setting parameters $k_1=1.2$ and $b=0.75$. For Doc2Query, we generate 40 synthetic queries per document using a pre-trained T5 (or mT5) model before indexing. For DeepCT, we employ the respective BERT-base models to project contextualized token embeddings into term weights.

\textbf{Dense and Personalized Retrieval.} For DPR, Sen-T5, MPNet, TEM, and CoPPS, we reproduce their results on this dataset using the
official open-source implementations and default training strategies. For personalized models (TEM and CoPPS), we explicitly feed the same historical interaction sequences (up to the exact same length as used in CHAP) into their sequence encoders to construct the user profile embeddings. Crucially, their dual-tower inner-product search is accelerated using the FAISS library to perform exhaustive vector search over the entire corpus. This ensures they are evaluated under an unbounded candidate budget, making the comparison strictly fair against CHAP, which operates under a restricted generative candidate pool.

\textbf{Generative Retrieval (GR) and Ranking Fairness.} To eliminate discrepancies caused by model scale and evaluate the intrinsic capability of each framework, all GR baselines (DSI, NCI, TIGER, LTRGR, MERGE, COBRA) are implemented using the same T5-base (or mT5-base) backbone. For ItemID construction in DSI and NCI, we employ Hierarchical K-Means on BERT embeddings to build semantic tree IDs. For TIGER and COBRA, we utilize their default RQ-VAE architecture configured with depth $L=3$ and codebook size $K=512$, strictly matching our CHAP's dimensionality.

To directly address potential concerns regarding evaluation fairness, particularly the candidate generation budgets: no external heavy rerankers (e.g., Cross-Encoders) are applied to any baseline or our CHAP model. During inference, we ensure a rigorously fair evaluation by restricting all GR models to the exact same candidate budget ($M=50$). 
Crucially, rather than enforcing an identical decoding algorithm that might artificially impair baseline performance (e.g., generating invalid IDs), we allow all GR baselines to employ their optimal Constrained Beam Search with a beam width of 50. In contrast, leveraging our decoupled architecture, CHAP efficiently generates 50 candidates via a Parallel Sampling strategy. 
For pure generative baselines, their final ranking relies exclusively on the beam search probabilities over their 50 candidates; for hybrid baselines (e.g., COBRA, MERGE), they utilize their internal scoring mechanisms over the same pool. Similarly, CHAP simply fuses its internally generated sparse probabilities and dense similarities in a parameter-free manner within its 50 samples. 

By aligning the language backbone, enforcing identical candidate budgets ($M=50$), maintaining the optimal decoding integrity of baselines, and restricting all models to internal parameter-free scoring, we guarantee that CHAP's superior ranking accuracy and remarkable inference throughput (QPS) stem entirely from our core architectural innovations.

\section{Ablation Variant Definitions}
\label{app:ablation_details}

In our ablation study (Sec. 5.3), we systematically deconstruct CHAP to validate the contribution of each module. The specific configurations of the ablated variants are defined as follows:

\textbf{(A) Hierarchical Semantic Alignment (HSA) Variants:}
\begin{itemize}
    \item \textit{w/o HSA}: Replaces our trainable, aligned query encoder with a standard RQ-VAE encoder whose weights are strictly frozen after the item-side pre-training stage.
    \item \textit{w/o $\mathcal{L}_\mathrm{CSA}$}: Excludes the cross-sample alignment losses (both Cross-Commitment and Cross-Reconstruction), relying only on contrastive and distillation objectives.
    \item \textit{w/o $\mathcal{L}_\mathrm{HCL}$}: Removes the hierarchical-aware contrastive learning objective, eliminating the explicit multi-granular alignment penalty.
    \item \textit{w/o $\mathcal{L}_\mathrm{Distill}$}: Omits the soft probability distillation module that regularizes the fine-tuned encoder's output distribution.
\end{itemize}

\textbf{(B) Sequence Modeling Variants:}
\begin{itemize}
    \item \textit{w/o Dense Refinement}: Eliminates the dense continuous vector from both the input sequence and the generation target. The model relies entirely on discrete sparse SIDs for sequence modeling and ranking.
\end{itemize}

\textbf{(C) Residual Cascading Generation Variants:}
\begin{itemize}
    \item \textit{w/o Residual Gen. (Autoregressive)}: Replaces the lightweight layer-wise residual fusion mechanism with a standard flat autoregressive generation, forcing the heavy Transformer Decoder to execute sequentially for all $L$ layers.
    \item \textit{w/o Self-Attention}: Retains the residual cascading mechanism but structurally removes all Masked Self-Attention sub-layers within the Transformer Decoder, retaining only the Cross-Attention mechanism to interact with the encoded history.
    \item \textit{w/o Decoder (Pooling + MLP)}: Entirely removes the Transformer Decoder. Instead, it relies on mean pooling over the encoder's outputs to derive a static global context vector, which is then fed directly into the residual blocks for SID and dense vector prediction.
\end{itemize}

\section{Definitions of Intent-wise Query Segments}
\label{app:segments}

In industrial Local-Life search scenarios, user intents are highly heterogeneous. To rigorously evaluate the model's robustness and the specific contribution of each architectural component, we extract four intent-wise query segments from the Local-Life dataset based on query attributes and historical interaction frequencies. Notably, to ensure a rigorous and realistic evaluation, the Ambiguous, ColdStart, and LongTail subsets share the exact same entire candidate item pool (i.e., over 10 million items) as the General set during retrieval. The detailed statistics of these segments are summarized in Tab.~\ref{tab:intent_segments_stat}.

\input{Sections/Tables/intent_wise_dataset_stat.tex}

\begin{itemize}
    \item \textbf{General:} This segment is exactly the entire set of the Local-Life dataset. It encompasses the natural flow of daily user search traffic, reflecting the macroscopic ranking ability of the model across all scenarios.
    \item \textbf{Ambiguous:} This segment comprises short, highly ambiguous queries with generalized intents (e.g., ``Dinner'', ``Nearby attractions'', ``Weekend entertainment''). These queries typically map to a vast number of potential candidate items across multiple categories. Excelling in this segment rigorously tests the model's \textit{Personalized Sequence Modeling} capability, as accurate retrieval strictly requires deducing explicit user preferences from historical behaviors to disambiguate the broad intent.
    \item \textbf{ColdStart:} This segment targets queries that exclusively lead to items introduced to the platform within the last 14 days. These items have extremely sparse or zero historical click logs. Standard generative models typically struggle on these data because they lack collaborative training signals for the new SIDs and fail to integrate dense representations for semantic fallback. This scenario directly evaluates the model's pure zero-shot semantic matching capability. 
    \item \textbf{LongTail:} This segment focuses on queries whose historical interaction frequencies fall into the bottom 10\% of the entire search logs (e.g., highly specific long-tail constraints, niche store names, or complex negated conditions). Similar to the ColdStart segment, standard GR models often exhibit poor performance on these data because they do not combine dense representations to capture precise textual semantics, falling back instead on memorized popularity bias. In contrast, excelling in this segment heavily relies on our model's intrinsic \textit{Hierarchical Semantic Alignment (HSA)} and dual-view design to accurately map complex linguistics.
\end{itemize}

\section{Additional Diagnostics for Semantic Alignment}
\label{app:alignment_diagnostics}

In addition to visualization, we use clustering-oriented quantitative diagnostics to measure whether the aligned representation space becomes more compact and separable. We consider Silhouette Score~\cite{rousseeuw1987silhouettes}, Calinski-Harabasz Index~\cite{calinski1974dendrite}, and Davies-Bouldin Index~\cite{davies1979cluster} over coarse product-category clusters constructed from the sampled encoder representations. Higher Silhouette and Calinski-Harabasz scores, together with a lower Davies-Bouldin score, indicate better intra-category compactness and inter-category separation. Tab.~\ref{tab:alignment_diagnostics} reports the corresponding metrics computed from 10,000 sampled encoder representations in ESCI-us using product-category labels as cluster assignments. 

\begin{table}[t]
\centering
\small
\resizebox{\linewidth}{!}{
\begin{tabular}{c|ccc}
\toprule
Method & Silhouette $\uparrow$ & CH Index $\uparrow$ & DB Index $\downarrow$ \\
\midrule
RQ-VAE & 0.218 & 612.4 & 1.684 \\
MERGE & 0.331 & 948.7 & 1.153 \\
CHAP & 0.387 & 1136.2 & 0.912 \\
\bottomrule
\end{tabular}
}
\caption{Semantic alignment diagnostics computed from 10,000 sampled encoder representations in ESCI-us.}
\label{tab:alignment_diagnostics}
\end{table}

\begin{figure*}[t]
    \centering
    \includegraphics[width=0.96\textwidth]{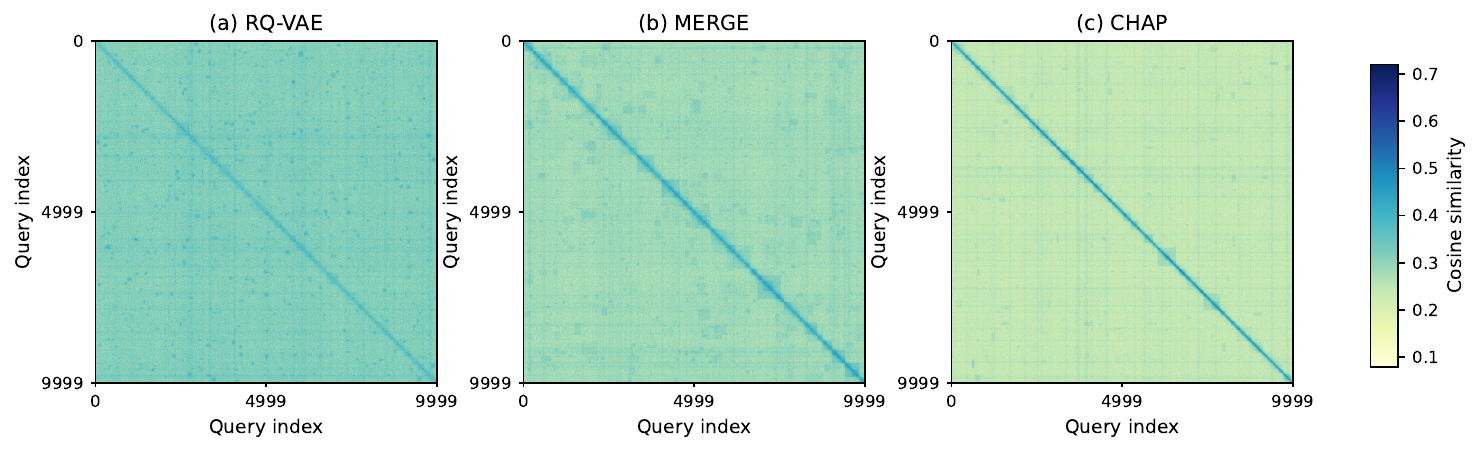}
    \caption{Similarity matrix diagnostics over 10,000 sampled queries from ESCI-us. The figure compares pairwise similarity matrices of RQ-VAE, MERGE, and CHAP, where CHAP exhibits a clearer local-neighborhood structure and fewer coarse diagonal blocks after semantic alignment.}
    \label{fig:alignment_similarity_matrix}
\end{figure*}

Following the matrix-style diagnostic in prior cascaded retrieval analysis, Fig.~\ref{fig:alignment_similarity_matrix} further provides a similarity-matrix view of the same trend. RQ-VAE shows a relatively noisy off-diagonal background, suggesting that item-only quantization leaves residual similarity between unrelated product categories. MERGE produces a clearer diagonal structure but still contains coarse diagonal blocks, indicating that its multi-level merging improves local consistency while retaining some coarse-grained semantic entanglement. CHAP exhibits a sharper local-neighborhood diagonal and a lighter off-diagonal background, suggesting that HSA better concentrates category-consistent representations while reducing spurious cross-category similarity.

\section{SID-Level Visualization Examples}
\label{app:sid_visualization_examples}

Following the qualitative SID analyses in MERGE~\cite{zhang2025multi}, we further inspect how CHAP organizes items across hierarchical SID layers. We use the real item-to-SID mappings from RQ-VAE, MERGE, and CHAP on ESCI-us, and select representative queries whose candidate items contain multiple relevance levels.

\begin{figure*}[t]
    \centering
    \includegraphics[width=0.96\textwidth]{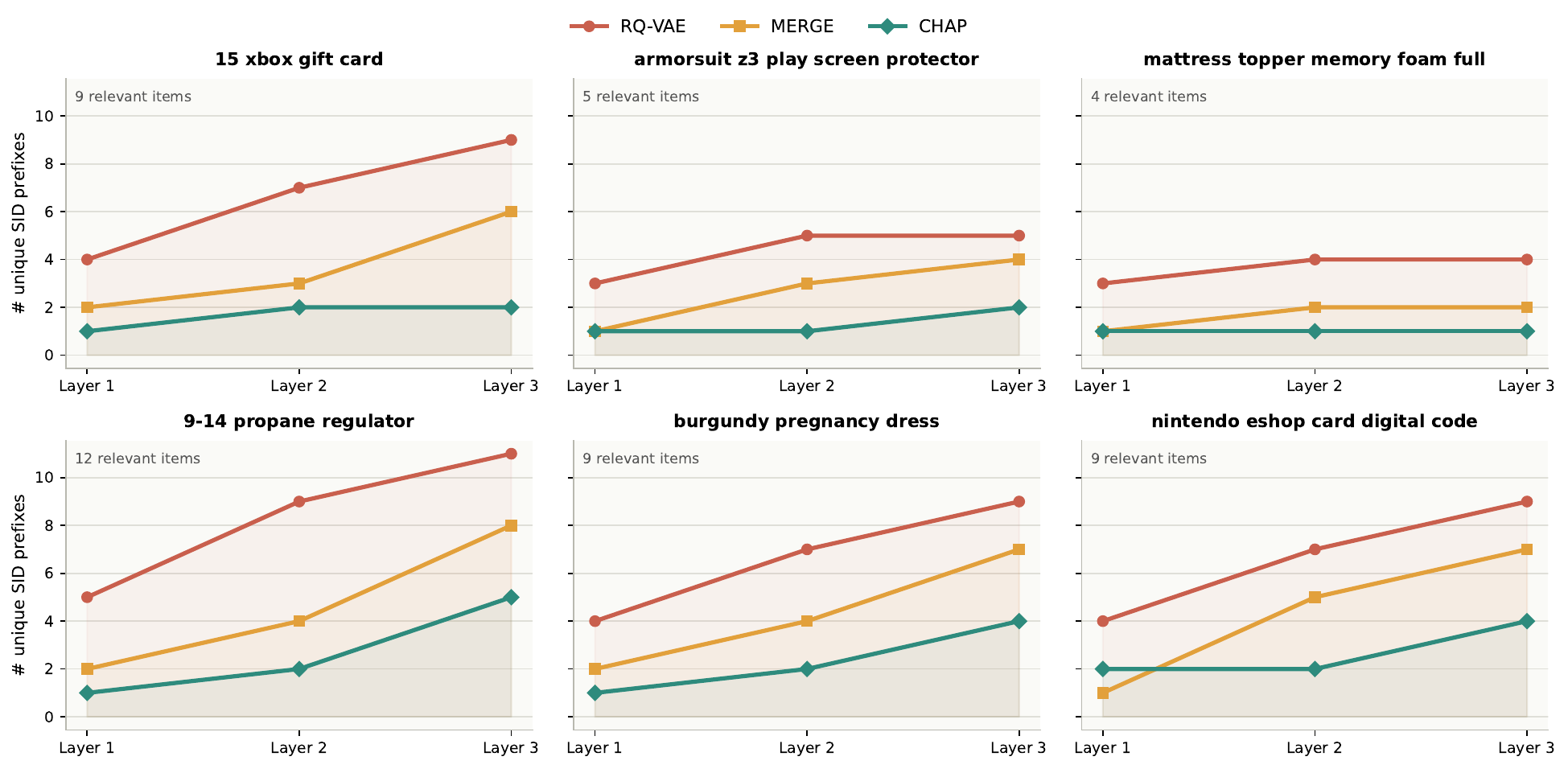}
    \caption{Layer-wise SID concentration over multiple representative ESCI-us queries. The curves are computed from the real item-to-SID mappings of RQ-VAE, MERGE, and CHAP. The y-axis reports the number of unique SID prefixes among query-relevant items, where lower values indicate more compact hierarchical assignments.}
    \label{fig:sid_layer_distribution_multiquery}
\end{figure*}

Fig.~\ref{fig:sid_layer_distribution_multiquery} visualizes the layer-wise SID concentration across six representative ESCI-us queries. For each query, we count the number of unique SID prefixes occupied by query-relevant items at each layer. Lower curves therefore indicate that relevant items are assigned to fewer hierarchical branches. Compared with RQ-VAE and MERGE, CHAP consistently produces more concentrated layer-wise distributions, especially at deeper SID layers. This trend suggests that HSA encourages the SID hierarchy to better follow query-conditioned relevance rather than only item-side lexical similarity.

\begin{figure*}[t]
    \centering
    \includegraphics[width=0.88\textwidth]{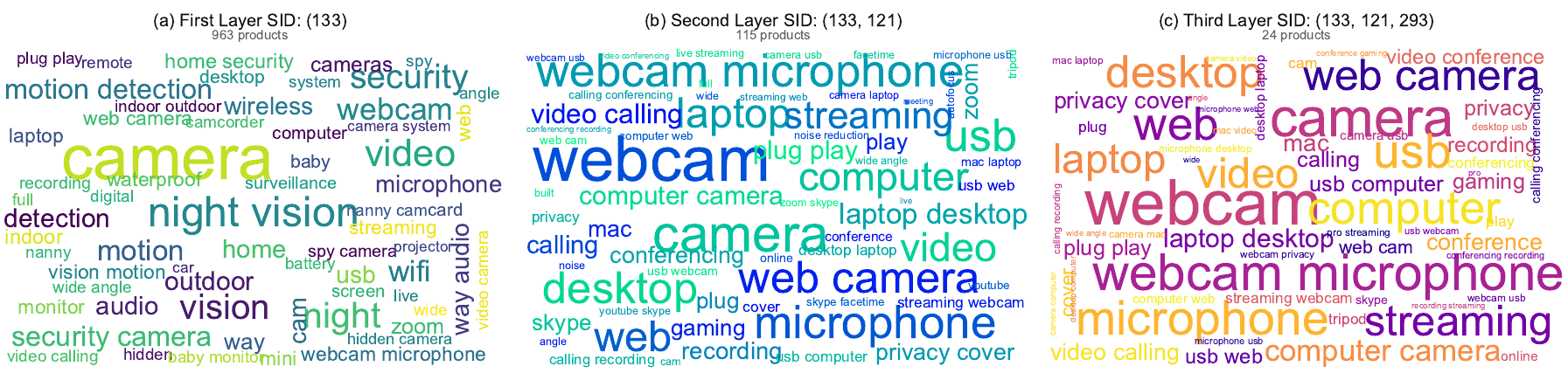}
    \caption{Product-title keyword-cloud visualization along an aligned camera-related SID path. The prefixes become progressively more specific from a broad camera-related cluster to webcam-oriented products and then to webcam products with microphone and streaming attributes.}
    \label{fig:sid_title_keywords_camera}
\end{figure*}

To examine the semantic meaning of SID layers, Fig.~\ref{fig:sid_title_keywords_camera} summarizes frequent product-title keywords along one aligned camera-related SID path. The first-layer prefix $(133)$ captures a coherent coarse-grained visual-device cluster, including camera, security camera, webcam, night vision, and motion detection products. The second-layer prefix $(133,121)$ narrows this cluster toward webcam and USB video-camera products, while the third-layer prefix $(133,121,293)$ further concentrates on webcam products with microphone, desktop, laptop, and streaming attributes. This coarse-to-fine keyword evolution is consistent with CHAP's hierarchical alignment objective.

Tab.~\ref{tab:sid_case_study_sandwich} provides item-level examples for a separate query, ``crustless sandwich maker''. Since CHAP is initialized from and fine-tuned upon the RQ-VAE codebook, the RQ-VAE and CHAP SID columns are comparable. In contrast, MERGE learns a separate codebook with its alignment loss, so its numeric SIDs should only be interpreted within MERGE's own codebook space rather than compared token by token with RQ-VAE or CHAP. In these selected examples, exact sandwich cutter and sealer products are mapped to the same CHAP SID, whereas irrelevant items such as generic cookie cutters, handheld sandwich grills, and craft makers remain separated. This example illustrates how aligned SIDs improve the consistency of generated ItemIDs for query-relevant items.

\begin{table*}[t]
\centering
\small
\resizebox{\textwidth}{!}{
\begin{tabular}{l|l|lll}
\toprule
Label & Product Title & RQ-VAE SID & MERGE SID & CHAP SID \\
\midrule
E & Sandwich Cutter and Sealer for Kids & $(14,127,331)$ & $(287,375,142)$ & $(14,127,294)$ \\
E & Decruster Sandwich Maker Set & $(14,502,331)$ & $(287,64,411)$ & $(14,127,331)$ \\
E & DIY Uncrustables Sandwich Cutters & $(160,127,393)$ & $(287,64,411)$ & $(14,127,331)$ \\
E & Fun Sandwich Cutters for Lunch Box & $(14,127,441)$ & $(287,64,156)$ & $(14,127,226)$ \\
I & Cookie Cutters with Irreverent Phrases & $(435,359,375)$ & $(93,382,27)$ & $(435,241,88)$ \\
I & Toas-Tite Handheld Sandwich Grill & $(472,154,266)$ & $(316,45,390)$ & $(472,308,119)$ \\
I & Cricut Champagne Maker & $(46,491,265)$ & $(446,220,108)$ & $(46,73,402)$ \\
\midrule
Q & crustless sandwich maker & $(14,127,393)$ & $(287,64,411)$ & $(14,127,331)$ \\
\bottomrule
\end{tabular}
}
\caption{Comparison of generated SIDs under the query ``crustless sandwich maker'' using RQ-VAE, MERGE, and CHAP.}
\label{tab:sid_case_study_sandwich}
\end{table*}

\section{AI Usage Disclosure}
\label{app:ai_usage}

In compliance with the conference guidelines regarding the use of artificial intelligence tools, we transparently disclose our light usage of AI assistants during the preparation of this manuscript. Specifically, AI tools were strictly limited to serving as utility aids for language polishing (e.g., correcting grammar, refining sentence structures, and improving readability for non-native speakers) and basic programming support (e.g., generating boilerplate scripts for data visualization). 

The AI assistants did not generate any novel scientific claims, experimental designs, or core methodology. All research conceptualization, model implementation, data analysis, and the intellectual contributions of this paper are exclusively the original work of the human authors.

%% file: Sections/Tables/open_source_dataset_stat_appendix.tex
\begin{table*}[t]
    \centering
    \caption{Detailed statistics of the four datasets.}
    \label{tab:open_source_dataset_stat_appendix}
    \resizebox{\textwidth}{!}{%
    \setlength\tabcolsep{6pt}
    \begin{tabular}{c|c c c|c c|c c|c}
        \toprule
        \multirow{2}{*}{\textbf{Dataset}} &
        \multirow{2}{*}{\textbf{\#Items}} &
        \multirow{2}{*}{\textbf{\#Queries}} &
        \multirow{2}{*}{\textbf{\#Q-I Pairs}} &
        \multicolumn{2}{c|}{\textbf{Train}} &
        \multicolumn{2}{c|}{\textbf{Test}} &
        \multirow{2}{*}{\textbf{Avg. Hist.}} \\
        \cmidrule(lr){5-6}\cmidrule(lr){7-8}
        & & & & \textbf{\#Queries} & \textbf{\#Q-I Pairs} & \textbf{\#Queries} & \textbf{\#Q-I Pairs} & \\
        \midrule
        ESCI-us    & 1,215,854  & 97,346     & 1,818,825  & 74,888  & 1,393,063 & 22,458 & 425,762 & -     \\
        KuaiSearch & 6,634,118  & 295,561    & 682,228    & 266,004 & 613,696   & 29,557 & 68,532  & 10.37 \\
        Amazon     & 35,772     & 9,070      & 9,070      & 6,896   & 6,896     & 2,174  & 2,174   & 40.12  \\
        Local-Life & 10,145,542 & 3,089,026 & 3,089,026 & 2,984,294 & 2,984,294 & 104,732 & 104,732 & 7.47  \\
        \bottomrule
    \end{tabular}%
    }
\end{table*}

%% file: Sections/Tables/intent_wise_dataset_stat.tex
\begin{table*}[t]
    \centering
    \caption{Detailed statistics of the intent-wise query segments extracted from the Local-Life dataset.}
    \label{tab:intent_segments_stat}
    \resizebox{\textwidth}{!}{%
    \setlength\tabcolsep{6pt}
    \begin{tabular}{c|c c c|c c|c c|c}
        \toprule
        \multirow{2}{*}{\textbf{Segment}} &
        \multirow{2}{*}{\textbf{\#Items}} &
        \multirow{2}{*}{\textbf{\#Queries}} &
        \multirow{2}{*}{\textbf{\#Q-I Pairs}} &
        \multicolumn{2}{c|}{\textbf{Train}} &
        \multicolumn{2}{c|}{\textbf{Test}} &
        \multirow{2}{*}{\textbf{Avg. Hist.}} \\
        \cmidrule(lr){5-6}\cmidrule(lr){7-8}
        & & & & \textbf{\#Queries} & \textbf{\#Q-I Pairs} & \textbf{\#Queries} & \textbf{\#Q-I Pairs} & \\
        \midrule
        General    & 10,145,542 & 3,089,026 & 3,089,026 & 2,984,294 & 2,984,294 & 104,732 & 104,732 & 7.47  \\
        Ambiguous  & 10,145,542 & 10,000   & 10,000   & -         & -         & 10,000  & 10,000  & 6.21  \\
        ColdStart  & 10,145,542 & 5,000   & 5,000   & -         & -         & 5,000  & 5,000  & 5.74  \\
        LongTail   & 10,145,542 & 10,000   & 10,000   & -         & -         & 10,000  & 10,000  & 10.16  \\
        \bottomrule
    \end{tabular}%
    }
\end{table*}